\documentclass[aps,prd,preprint,superscriptaddress,tightenlines,nofootinbib]{revtex4-1}

\usepackage{graphicx}% Include figure files
\usepackage{dcolumn}% Align table columns on decimal point
\usepackage{bm}% bold math
\usepackage{epsfig}
\usepackage{amsmath}

\begin{document}
%===================> ADD here your LATEX definitions

\newcommand{\ee}{e$^+$e$^-$}
\newcommand{\ff}{f$_{2}$(1525)}
\newcommand{\bb}{$b \overline{b}$}
\newcommand{\cc}{$c \overline{c}$}
\newcommand{\sbs}{$s \overline{s}$}
\newcommand{\uu}{$u \overline{u}$}
\newcommand{\dd}{$d \overline{d}$}
\newcommand{\qq}{$q \overline{q}$}
\newcommand{\suo}{\rm{\mbox{$\epsilon_{b}$}}}
\newcommand{\loro}{\rm{\mbox{$\epsilon_{c}$}}}
\newcommand{\kos}{\ifmmode \mathrm{K^{0}_{S}} \else K$^{0}_{\mathrm S} $ \fi}
\newcommand{\kol}{\ifmmode \mathrm{K^{0}_{L}} \else K$^{0}_{\mathrm L} $ \fi}
\newcommand{\ko}{\ifmmode {\mathrm K^{0}} \else K$^{0} $ \fi}

\def\tpc{three-particle correlation}
\def\twopc{two-particle correlation}
\def\ksks{K$^0_S$K$^0_S$}
\def\ee{e$^+$e$^-$}
\def\ff{f$_{2}$(1525)}

\hfill{CLNS 10/2065}

\hfill{CLEO 10-03}

\title{Study of $\bm{\psi(2S)}$ Decays to $\bm{\gamma p\bar{p}}$, $\bm{\pi^0 p\bar{p}}$ and $\bm{\eta p\bar{p}}$ and Search for $\bm{p\bar{p}}$ Threshold Enhancements}

\author{J.~P.~Alexander}
\author{D.~G.~Cassel}
\author{S.~Das}
\author{R.~Ehrlich}
\author{L.~Fields}
\author{L.~Gibbons}
\author{S.~W.~Gray}
\author{D.~L.~Hartill}
\author{B.~K.~Heltsley}
\author{D.~L.~Kreinick}
\author{V.~E.~Kuznetsov}
\author{J.~R.~Patterson}
\author{D.~Peterson}
\author{D.~Riley}
\author{A.~Ryd}
\author{A.~J.~Sadoff}
\author{X.~Shi}
\author{W.~M.~Sun}
\affiliation{Cornell University, Ithaca, New York 14853, USA}
\author{J.~Yelton}
\affiliation{University of Florida, Gainesville, Florida 32611, USA}
\author{P.~Rubin}
\affiliation{George Mason University, Fairfax, Virginia 22030, USA}
\author{N.~Lowrey}
\author{S.~Mehrabyan}
\author{M.~Selen}
\author{J.~Wiss}
\affiliation{University of Illinois, Urbana-Champaign, Illinois 61801, USA}
\author{J.~Libby}
\affiliation{Indian Institute of Technology Madras, Chennai, Tamil Nadu 600036, India}
\author{M.~Kornicer}
\author{R.~E.~Mitchell}
\author{M.~R.~Shepherd}
\author{C.~M.~Tarbert}
\affiliation{Indiana University, Bloomington, Indiana 47405, USA }
\author{D.~Besson}
\affiliation{University of Kansas, Lawrence, Kansas 66045, USA}
\author{T.~K.~Pedlar}
\author{J.~Xavier}
\affiliation{Luther College, Decorah, Iowa 52101, USA}
\author{D.~Cronin-Hennessy}
\author{J.~Hietala}
\author{P.~Zweber}
\affiliation{University of Minnesota, Minneapolis, Minnesota 55455, USA}
\author{S.~Dobbs}
\author{Z.~Metreveli}
\author{K.~K.~Seth}
\author{A.~Tomaradze}
\author{T.~Xiao}
\affiliation{Northwestern University, Evanston, Illinois 60208, USA}
\author{S.~Brisbane}
\author{L.~Martin}
\author{A.~Powell}
\author{P.~Spradlin}
\author{G.~Wilkinson}
\affiliation{University of Oxford, Oxford OX1 3RH, UK}
\author{H.~Mendez}
\affiliation{University of Puerto Rico, Mayaguez, Puerto Rico 00681}
\author{J.~Y.~Ge}
\author{D.~H.~Miller}
\author{I.~P.~J.~Shipsey}
\author{B.~Xin}
\affiliation{Purdue University, West Lafayette, Indiana 47907, USA}
\author{G.~S.~Adams}
\author{D.~Hu}
\author{B.~Moziak}
\author{J.~Napolitano}
\affiliation{Rensselaer Polytechnic Institute, Troy, New York 12180, USA}
\author{K.~M.~Ecklund}
\affiliation{Rice University, Houston, Texas 77005, USA}
\author{J.~Insler}
\author{H.~Muramatsu}
\author{C.~S.~Park}
\author{L.~J.~Pearson}
\author{E.~H.~Thorndike}
\author{F.~Yang}
\affiliation{University of Rochester, Rochester, New York 14627, USA}
\author{S.~Ricciardi}
\affiliation{STFC Rutherford Appleton Laboratory, Chilton, Didcot, Oxfordshire, OX11 0QX, UK}
\author{C.~Thomas}
\affiliation{University of Oxford, Oxford OX1 3RH, UK}
\affiliation{STFC Rutherford Appleton Laboratory, Chilton, Didcot, Oxfordshire, OX11 0QX, UK}
\author{M.~Artuso}
\author{S.~Blusk}
\author{R.~Mountain}
\author{T.~Skwarnicki}
\author{S.~Stone}
\author{J.~C.~Wang}
\author{L.~M.~Zhang}
\affiliation{Syracuse University, Syracuse, New York 13244, USA}
\author{G.~Bonvicini}
\author{D.~Cinabro}
\author{A.~Lincoln}
\author{M.~J.~Smith}
\author{P.~Zhou}
\author{J.~Zhu}
\affiliation{Wayne State University, Detroit, Michigan 48202, USA}
\author{P.~Naik}
\author{J.~Rademacker}
\affiliation{University of Bristol, Bristol BS8 1TL, UK}
\author{D.~M.~Asner}
\altaffiliation[Now at: ]{Pacific Northwest National Laboratory, Richland, WA 99352}
\author{K.~W.~Edwards}
\author{K.~Randrianarivony}
\author{G.~Tatishvili}
\altaffiliation[Now at: ]{Pacific Northwest National Laboratory, Richland, WA 99352}
\affiliation{Carleton University, Ottawa, Ontario, Canada K1S 5B6}
\author{R.~A.~Briere}
\author{H.~Vogel}
\affiliation{Carnegie Mellon University, Pittsburgh, Pennsylvania 15213, USA}
\author{P.~U.~E.~Onyisi}
\author{J.~L.~Rosner}
\affiliation{University of Chicago, Chicago, Illinois 60637, USA}
\collaboration{CLEO Collaboration}
\noaffiliation

%-------- END INSERT ------------

\date{\today}

\begin{abstract}
The decays of $\psi(2S)$ into $\gamma p\bar{p}$, $\pi^{0}p\bar{p}$ and $\eta p\bar{p}$
have been studied with the CLEO-c detector using a sample of 24.5 million $\psi(2S)$ events
obtained from $e^{+}e^{-}$ annihilations at $\sqrt{s}$ = 3686 MeV. The data show evidence 
for the excitation of several $N^{*}$ resonances
in $p\pi^{0}$ and $p\eta$ channels in $\pi^{0}p\bar{p}$ and $\eta p\bar{p}$ decays, and 
$f_{2}$ states in $\gamma p\bar{p}$ decay. Branching fractions for decays of $\psi(2S)$ 
to $\gamma p\bar{p}$, $\pi^{0} p\bar{p}$ and $\eta p\bar{p}$ have been determined.
No evidence for $p\bar{p}$ threshold enhancements was found in the reactions 
$\psi(2S)\to X p\bar{p}$, where X = $\gamma$,$\pi^{0}$,$\eta$. 
We do, however, find confirming evidence for a $p\bar{p}$ threshold enhancement in
$J/\psi \to \gamma p\bar{p}$  as previously reported by BES.

\end{abstract}

\pacs{14.40Gx, 13.25Gv,13.66Bx}
\maketitle

\section{Introduction}

There is long-standing interest in 6-quark dibaryons and 3 quark - 3 antiquark 
``baryonium'' states which are permitted in QCD, and may possibly exist. Of particular
interest is a possible bound state of a proton and an antiproton. The $p\bar{p}$
state, sometimes called ``protonium'', was searched for in many experiments,
but no credible evidence was found~\cite{prot1,prot2}. Interest was revived in 2002 by two
reports by the Belle Collaboration of threshold enhancements in $M(p\bar{p})$ in
the decays $B^{\pm} \to K^{\pm}p\bar{p}$~\cite{belle1} and $\bar{B}^{0}\to D^{*0}p\bar{p}$~\cite{belle2}.
These reports were followed by a BES report of threshold enhancement in the decay
$J/\psi \to \gamma p\bar{p}$~\cite{bes}. Subsequently, there have been reports of threshold
enhancements and studies by Belle in $B^{+}\to \pi^{+}p\bar{p}$, 
$B^{0}\to K^{0}p\bar{p}$, and $B^{+}\to K^{*+}p\bar{p}$~\cite{belle3}; 
by BaBar in $B^{+}\to K^{+}p\bar{p}$~\cite{babar1}; and
$B^{0} \to p\bar{p}+(\bar{D}^{0},~\bar{D}^{*0},~D^{-}\pi^{+},
~\mathrm{or}~D^{*-}\pi^{+})$~\cite{babar2}; and
most recently by Belle in $B^{+}\to K^{+}p\bar{p}$ and 
$B^{+}\to \pi^{+}p\bar{p}$~\cite{belle4}.
Many theoretical explanations, cusp effects, final state interactions, quark fragmentation,
and real bound states of quarks and gluons, have been suggested for these threshold
enhancements~\cite{threv}.

If the enhancement reported by BES in the decay $J/\psi \to \gamma p\bar{p}$~\cite{bes} is 
due to a threshold resonance, it is reasonable to expect that evidence for it may be found 
also in $\psi(2S) \to \gamma p\bar{p}$. Further insight into its nature may be provided 
by the study of the reactions $\psi(2S) \to \pi^{0}p\bar{p}$ and $\eta p\bar{p}$.

\section{Event Selection}

In this paper we report on studies of these reactions observed in the CLEO-c detector
in a data sample of 24.5 million $\psi(2S)$ events obtained by $e^{+}e^{-}$
annihilations at $\sqrt{s}=$3.686 GeV at the Cornell Electron Storage Ring, CESR.
In addition, we use 20.7 pb$^{-1}$ of off-resonance data taken at $\sqrt{s}=3.67$ GeV.

The CLEO-c detector, described in detail elsewhere~\cite{cleodet}, has a solid angle 
coverage of 93$\%$ for charged and neutral particles. The charged particle tracking and 
identification system 
operates in a 1.0 T solenoidal magnetic field, and consists of an inner drift
chamber, a central drift chamber, and a ring-imaging Cherenkov (RICH) detector.
It has a momentum resolution of $\sim$0.6$\%$ at momenta of $\sim$1 GeV/$c$.
The CsI elecromagnetic calorimeter has a photon energy resolution of $\sim$2.2$\%$
for $E_{\gamma}=1$ GeV and $\sim$5$\%$ at 100 MeV.

Photons and charged particles with $|\cos \theta |<0.8$ were accepted in the detector, 
where $\theta$ is the polar angle with respect to the incoming positron beam. 
For the modes involving the direct decays of the $\psi(2S)$, exactly two oppositely 
charged tracks were required in candidate events. 
A photon candidate was defined as a shower which does not match a track within 100
mrad, is not in one of the few cells of the electromagnetic calorimeter known to be noisy, has the transverse
distribution of energy consistent with an electromagnetic shower, and has an energy
more than 30 MeV.
For $\gamma p\bar{p}$ the number of showers was required to be $\geq 1$, and it was
required that the most energetic shower (the signal photon candidate) does not make a $\pi^{0}$
or $\eta$ with any other shower with a pull mass $<3\sigma$. For 
$\pi^{0}p\bar{p}$ and $\eta p\bar{p}$ the number of showers was required to be $\geq 2$. 

To identify charged tracks as protons and antiprotons, the energy loss in the 
drift chambers ($dE/dx$) and RICH information was used.
For tracks of momentum less than 0.9 GeV/$c$, only $dE/dx$ information is used.
To utilize $dE/dx$ information, for each particle hypothesis, $X=\pi,~K,~p$ or $\bar{p}$,
we calculate $\chi_{X}^{dE/dx}=[(dE/dx)_\mathrm{meas}-(dE/dx)_\mathrm{pred}]/\sigma_{X}$, where 
$(dE/dx)_\mathrm{meas}$ is the measured value of $dE/dx$, $(dE/dx)_\mathrm{pred}$ is 
the predicted value for hypothesis $X$, and $\sigma_{X}$ is the standard deviation of 
the measurements for hypothesis $X$. We cut on both the 
deviation of the measured $dE/dx$ from a given particle hypothesis, 
$\chi_X^{dE/dx}$, and the difference in $\chi^{dE/dx}$ 
between two particle hypotheses, 
$\Delta \chi_{X,Y}^{dE/dx} \equiv \chi_X^{dE/dx} - \chi_Y^{dE/dx}$.  
For higher momentum tracks, we use a combined log-likelihood variable. 
For example, to differentiate between proton and pion we construct
$$\Delta \mathcal{L}_{p,\pi} = (\chi_p^{dE/dx})^2 - (\chi_\pi^{dE/dx})^2 + 2\times (L_p^{RICH} - 
L_\pi^{RICH})$$
where $L_{\pi,p}^{RICH}$ are the log-likelihoods obtained from the RICH subdetector.  
We use RICH information if the track has $|\cos \theta|<0.8$ and the track has 
valid RICH information for at least one hypothesis (pion or proton), and at least three
photons consistent with that hypothesis were recorded in the RICH.

We consider three different momentum regions for charged tracks.
\begin{itemize}
\item \textbf{p $\bm{<}$ 0.9 GeV/$\bm{c}$:} In this momentum region only $dE/dx$ information 
for the tracks is available, and it is required that it be within
$3\sigma_{p}$ of the proton hypothesis, and must be more ``proton-like'' than 
``pion-like'' or ``kaon-like'', {\it i.e.} $|\chi_p^{dE/dx}| < 3$, 
$\Delta \chi_{\pi,p}>0$, and $\Delta \chi_{K,p}>0$.

\item \textbf{0.9 GeV/$\bm{c}$ $\bm{<}$ p $\bm{<}$ 1.15 GeV/$\bm{c}$:} In this momentum region, 
although we are above the threshold for a proton to emit Cherenkov radiation 
in the RICH, the probability that it will do so is still low.  Therefore, if 
RICH information is available, we require that the track be more ``proton-like'' 
in the combined log-likelihood variable, {\it i.e.}, $\Delta \mathcal{L}_{p,\pi} < 0$.  
If RICH information is not available, we again require that 
$|\chi_p^{dE/dx}| < 3$, and additionally require a $5\sigma$ difference between 
the proton hypothesis and the pion and kaon hypotheses, {\it i.e.}, 
$\Delta \chi_{\pi,p}>5$, and $\Delta \chi_{K,p}>5$, in order to reduce the 
number of other particles which pass these cuts.

\item \textbf{p $\bm{>}$ 1.15 GeV/$\bm{c}$:} In this momentum region, $dE/dx$ alone no longer 
provides useful information for proton identification.  We require that RICH 
information be available, and that $\Delta \mathcal{L}_{p,\pi} < 0$.
\end{itemize}

We require one of the charged tracks to be identified as a proton or antiproton 
and assume the other track to be its antiparticle as required by baryon conservation,
and we require the proton and antiproton to come from a common vertex, with kinematic fit yielding $\chi^{2}_{p\bar{p}~\mathrm{vertex}}<20$.

Finally, in order to select the events for the channels of interest: 
\begin{itemize}
\item For selection of $\psi(2S)\to \gamma p\bar{p}$ events we require 
$\chi^{2}_{\mathrm{fit}}/\mathrm{degrees~of~freedom~(d.o.f.)}<5$ for the four-momentum 
conservation constrained fit to $p$, $\bar{p}$ and the signal photon candidate.

\item For selection of $\psi(2S) \to \pi^{0}(\eta)p\bar{p}$ events, we first require
that only one $\pi^{0}$($\eta$) be made by any two photons and the pull mass
be within 3$\sigma$. Then we require $\chi^{2}_{\mathrm{fit}}/\mathrm{d.o.f.}<5$
for the four-momentum conservation constrained fit to $p$, $\bar{p}$, and 
$\pi^{0}(\eta)$. We remove the events corresponding to $\psi(2S)\to \pi^{0}(\eta) J/\psi$
by rejecting candidates for which $M(p\bar{p})=M(J/\psi)\pm 20$ MeV/$c^{2}$. Figure~1 
shows the distribution of $M(\gamma \gamma)$ before and after the selection of 
$\pi^{0}$ and $\eta$ described above.

\item For selection of $\psi(2S)\to \pi^{+}\pi^{-}J/\psi$, $J/\psi \to \gamma p\bar{p}$
events the additional event selection requirements are described in Sec.~VIII. 
\end{itemize}

The values of the $\chi^{2}$ cuts for the fits were selected based on the comparison 
of the data and the phase space distributions for the individual decays obtained
from Monte Carlo (MC) simulations.

\begin{figure}[ht!]
\begin{center}
\includegraphics[width=3.2in]{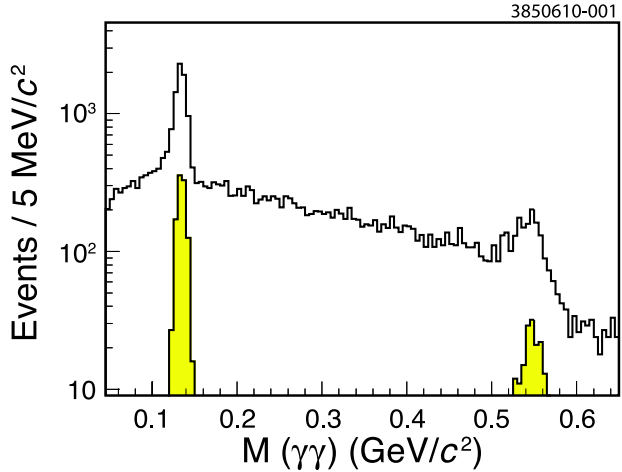}
\end{center}
\caption{
Distribution of $M(\gamma \gamma)$ in $\psi(2S)\to \pi^{0}(\eta)p\bar{p}$.
The unshaded histogram shows the $M(\gamma \gamma)$ distribution before the selection of
$\pi^{0}$ and $\eta$, and the shaded histogram shows it after the $\pi^{0}$ and $\eta$
selection described in the text.}
\end{figure}

\section{Monte Carlo Studies}

In order to verify the event selection criteria and determine efficiencies, 
50,000 phase space MC events were generated for each decay channel 
analyzed. As an example, for $\psi(2S)\to \gamma p\bar{p}$ events with 
$M(p\bar{p})<$2.85 GeV/$c^{2}$ the contribution of each step of event selection is presented 
in Table~I. 
The overall phase space efficiency is $(27.7\pm0.2)\%$. The corresponding efficiency for 
$\psi(2S)\to \pi^{0}p\bar{p}$ is $(26.9\pm0.2)\%$, and for $\psi(2S)\to \eta p\bar{p}$ 
it is $(27.7\pm0.2)\%$. 

\begin{table}[h!]
\begin{center}
\caption{Efficiencies of the individual event selection criteria for the decay
$\psi(2S)\to \gamma p\bar{p}$ based on phase space MC simulation.}
\begin{tabular}{lc}
\hline \hline
Selection requirement & Efficiency ($\%$) \\ \hline
Charged track and photon selection & \\
$N_{ch}$ = 2, net charge = 0, $N_{\gamma}\geq$1 & 74.5 \\ 
& \\
Signal photon does not make &  \\
$\pi^{0}$ with any other shower & 97.0 \\ 
& \\
Vertex fit, constrained fit & \\
$\chi^{2}_\mathrm{vertex}<20$, $\chi^{2}_\mathrm{fit}<5$ & 84.3 \\
& \\
Proton-antiproton identification & 98.2 \\ 
& \\
$M(p\bar{p})<$2.85 GeV/$c^{2}$  & 64.3 \\ 
& \\
All $p$, $\bar{p}$ and most energetic photon & \\
are in the barrel ($|\cos \theta| < 0.8$) & 72.0 \\ \hline 
Total & 27.7 \\ \hline \hline
\end{tabular}
\end{center}
\end{table}

\begin{figure*}[tbh!]
\begin{center}
\includegraphics[width=6.4in]{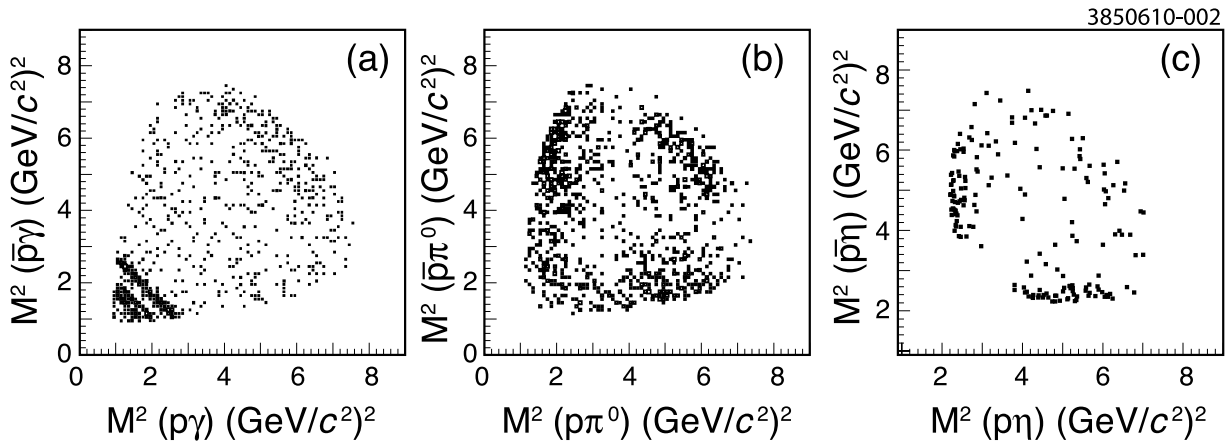}
\end{center}
\caption{(a) Dalitz plots for the data: (a) $M^{2}(p\gamma)$ versus
$M^{2}(\bar{p}\gamma)$
for $\psi(2S)\to \gamma p\bar{p}$; (b) $M^{2}(p\pi^{0})$ versus $M^{2}(\bar{p}\pi^{0})$
for $\psi(2S)\to \pi^{0}p\bar{p}$; (c) $M^{2}(p\eta)$ versus $M^{2}(\bar{p}\eta)$
for $\psi(2S)\to \eta p\bar{p}$.}
\end{figure*}

We also use $\psi(2S)$ ``generic'' MC events with the
available statistics of about five times data events ($\sim$118 million events).
The generic $\psi(2S)$ MC sample is generated using the available branching fractions
for the $\psi(2S)$, $\chi_{cJ}$, $J/\psi$, and $\eta_{c}$ decays, with unmeasured
decay modes simulated by JETSET~\cite{jetset}.
We have tested the event selection using a generic MC sample. 
We apply the same event selection to these MC events, extract the different 
branching fractions, and compare them to the branching fractions which were input in creating the
generic MC sample. The agreement between the input and output branching fractions
for $\psi(2S)\to \gamma p\bar{p}$, $\pi^{0}p\bar{p}$, and $\eta p\bar{p}$, is found
to be within $(2.4\pm 3.9)\%$, $(1.0\pm 1.0)\%$, and $(3.0\pm 4.0)\%$, respectively.

\section{Overview of $\bm{\psi(2S)}$ Decays}

Figure~2 shows Dalitz plots for the data for the three decays. All three 
Dalitz plots show event populations which are far from uniform, as would be expected for pure 
phase space decays, and suggest
contribution by intermediate excited nucleon states, $N^{*}$, and mesons.
Since the branching fractions for $N^{*}$ decays to $N\pi$ and $N\eta$ are generally 
much larger than those for decays to $N\gamma$, we expect excitation of $N^{*}$ states in 
$\psi(2S)\to \pi^{0} p\bar{p}$ and $\psi(2S)\to \eta p\bar{p}$. Similarly, we expect
excitation of intermediate meson states which decay into $p\bar{p}$, $f_{J}$ states in 
$\psi(2S)\to \gamma p\bar{p}$ and $\eta p\bar{p}$, and $a_{J}$ states in 
$\psi(2S)\to \pi^{0}p\bar{p}$.

A caveat about the intermediate states is in order.
The intermediate $N^{*}$ and $f_{0,2}$ states which we use in our analysis tend to have
masses $\geq 1.5$ GeV/$c^{2}$ . Unfortunately, the existence of most high mass $N^{*}$ 
resonances is either uncertain or poorly established, and their masses and widths have 
large uncertainties, so much so that the 2008 Particle Data review (PDG08)~\cite{pdg08} 
omits many of them from its summary table. 
Similar uncertainties exist for meson states with masses $\geq 1.5$ GeV/$c^{2}$ . 
Therefore our identification of an observed resonance with a known resonance is 
necessarily tentative.

Our data lack the statistics to make a full partial wave analysis. Instead, we analyse
the projections of the Dalitz plots of invariant mass distributions for 
$M(p(\gamma,\pi^{0},\eta))$ and $M(p\bar{p})$. Throughout this paper, charge conjugate 
states and their contributions are implied. 

We fit the invariant mass distributions with contributions from phase space and the
minimum number of resonances required to obtain good fits. The resonances are parametrized
in terms of relativistic Breit-Wigner functions with mass dependent widths and include
the Blatt-Weisskopf penetration factors [13, see p.~772]. We note that peak 
positions and 
widths in the relativistic fits can be substantially different from those for the simple
Breit-Wigner function, particularly for large widths and proximity to 
thresholds~\cite{pdg08}. 

In order to take proper account of intermediate states and possible reflections
in the Dalitz plots we analyze the data in the full range of 
$p\bar{p}$ invariant mass, from threshold to 3.6 GeV/$c^{2}$ . In this important respect the present
analysis differs from the BES analyses~\cite{bes,besrad,bespsip}.

In the following Sections, V, VI and VII we discuss the decays
$\psi(2S)\to \gamma p\bar{p}$, $\pi^{0} p\bar{p}$ and $\eta p\bar{p}$,
respectively. In Sec. VIII we present the results of the analysis of our limited statistics
sample of $J/\psi \to \gamma p\bar{p}$ events.

\section{The Decay $\bm{\psi(2S) \to \gamma p\bar{p}}$}

For the search for a threshold enhancement in $\psi(2S)\to \gamma p\bar{p}$, and
for the measurement of the inclusive branching fraction 
$\mathcal{B}(\psi(2S) \to \gamma p\bar{p})$ we limit ourselves to
$M(p\bar{p})<2.85$ GeV/$c^{2}$, below the $\eta_{c}$ mass. 
As a check on our analysis technique, we use our data for 
$M(p\bar{p})>3.15$ GeV/$c^{2}$  to calculate the $\mathcal{B}(\chi_{cJ} \to p\bar{p})$ 
branching fractions and compare them to recent measurements.

\begin{figure}[th!]
\begin{center}
\includegraphics[width=4.5in]{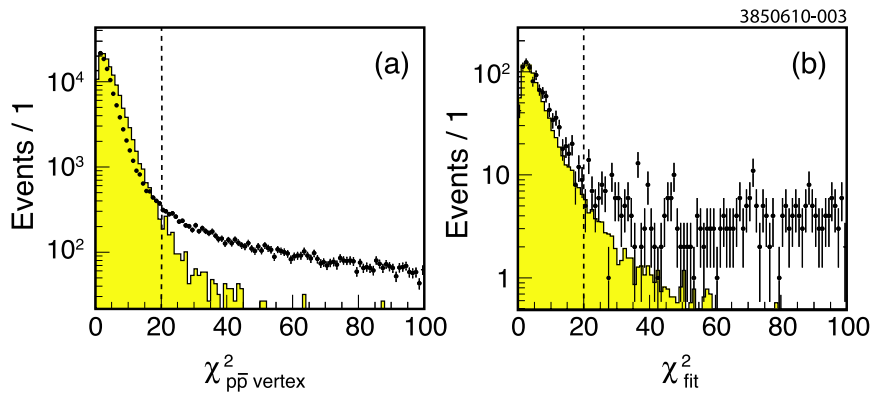}
\end{center}
\caption{
Distributions of the $\chi^{2}$ of four-momentum conservation constrained fits
for $\psi(2S)\to \gamma p\bar{p}$: (a) $\chi^{2}$ of vertex fit,
(b) $\chi^{2}$ of the four-momentum conservation fit. Points correspond
to the data and the shaded histograms correspond to the phase space MC simulation.
Dashed lines indicate the cut values used.
}
\end{figure}

\begin{figure}[th!]
\begin{center}
\includegraphics[width=4.5in]{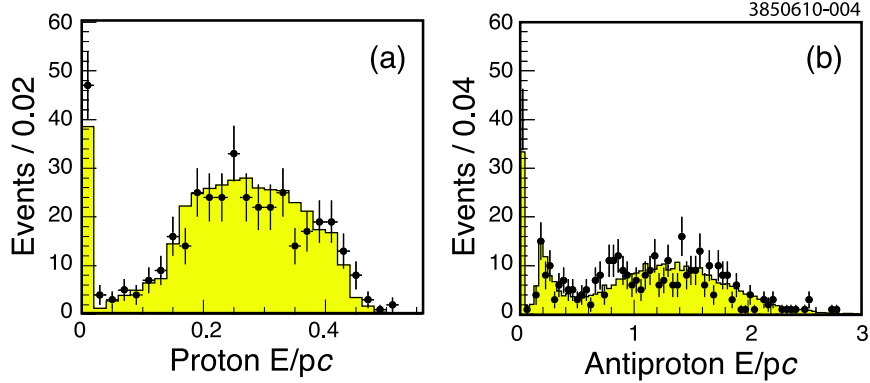}
\end{center}
\caption{
Distributions of E/p$c$ for $\psi(2S)\to \gamma p\bar{p}$ for 
$M(p\bar{p})<2.85$ GeV/$c^{2}$:
(a) protons, (b) antiprotons. Because of annihilations
E/p for antiprotons extends over a much larger range than for protons.
Points correspond to the data and the shaded histograms to the phase space MC
simulation.}
\end{figure}

\begin{figure}[th!]
\begin{center}
\includegraphics[width=4.5in]{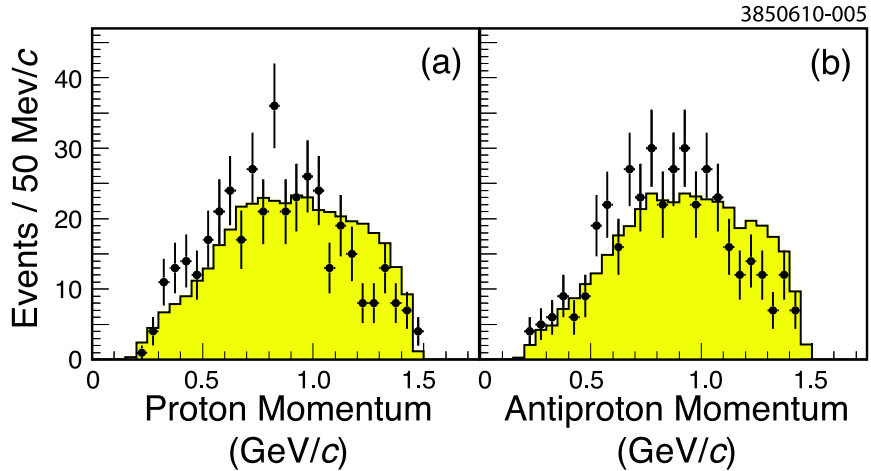}
\end{center}
\caption{
Distributions of the momenta of charged particles for
$\psi(2S)\to \gamma p\bar{p}$: (a) protons, (b) antiprotons.
Points correspond to the data and the shaded histograms to the phase space 
MC simulation.
}
\end{figure}

\begin{table*}[tb!]
\begin{center}
\caption{Results for $\mathcal{B}(\chi_{cJ}\to p\bar{p})$ for $\chi_{cJ}$ states.
The PDG08 values for $\mathcal{B}(\psi(2S)\to \gamma \chi_{cJ})$ have been
used to obtain the results for $\mathcal{B}(\chi_{cJ}\to p\bar{p})$.
}
\begin{tabular}{llll}
\hline \hline
& $\chi_{c0}$ & $\chi_{c1}$ & $\chi_{c2}$ \\
\hline
Mass (MeV/c$^{2}$) & 3412.8$\pm$1.0 & 3512.5$\pm$0.4 & 3555.0$\pm$1.0 \\
N(events) & 236.0 $\pm$ 18.4 & 79.0 $\pm$ 10.7 & 62.5 $\pm$ 9.8 \\
Efficiency in $\%$ & 39.5 & 41.4 & 37.9 \\
$\mathcal{B}(\psi(2S)\to \gamma \chi_{cJ})\times \mathcal{B}(\chi_{cJ}\to p\bar{p})
\times 10^{6}$ & 22.0 $\pm$ 1.7 & 7.05 $\pm$ 0.96 & 6.07 $\pm$ 0.95 \\
$\mathcal{B}(\chi_{cJ}\to p\bar{p})\times 10^{5}$ (this analysis) & 23.4 $\pm$
2.1 & 8.0 $\pm$ 1.1 & 7.3 $\pm$ 1.2 \\
$\mathcal{B}(\chi_{cJ}\to p\bar{p})\times 10^{5}$ (CLEO~\cite{cleopp}) & 25.7
$\pm$ 1.5 $\pm$ 2.0 & 9.0 $\pm$ 0.8 $\pm$ 0.6 & 7.7 $\pm$ 0.8 $\pm$ 0.6 \\
$\mathcal{B}(\chi_{cJ}\to p\bar{p})\times 10^{5}$ (PDG08~\cite{pdg08}) & 21.4
$\pm$ 1.9 & 6.6 $\pm$ 0.5 & 6.7 $\pm$ 0.5 \\
\hline \hline
\end{tabular}
\end{center}
\end{table*}

In Fig.~3 we show the $\chi^{2}$ distributions for the data and the phase space 
MC events for vertex fit and four-momentum conservation constrained fit to the proton, 
antiproton and most energetic shower in the event. All other selection criteria 
have been applied. Comparison of these distributions suggests the cut values 
$\chi^{2}_\mathrm{p\bar{p}~vertex}<20$ and $\chi^{2}_\mathrm{fit}<20$.

In Fig.~4 we compare E/p$c$ distributions for protons and antiprotons in data 
and the phase space $\psi(2S)\to \gamma p\bar{p}$ MC simulation, 
where E is the energy determined from the calorimeter and p is the momentum determined 
from track reconstruction. The distributions of 
protons and antiprotons are different, because antiprotons annihilate in the 
material of the electromagnetic calorimeter. However, for both protons and 
antiprotons, the data and the phase space MC distributions show good 
agreement.

In Fig.~5 proton and antiproton momentum distributions for the data and the
phase space $\psi(2S)\to \gamma p\bar{p}$ MC simulation are shown. The agreement between 
MC simulation and data momentum distributions is not good, and may indicate the effect 
of intermediate resonances.

In Fig.~6 we present the $M(p\bar{p})$ invariant mass distributions for the data and
the $\psi(2S)$ generic MC simulation, which includes the excitation of $\chi_{cJ}$ and
$\eta_{c}$, but not the ISR generated $J/\psi$. All event selection criteria 
have been applied. The generic MC events are normalized to the number of 
$\psi(2S)$ events in the data for a qualitative comparison.

\begin{figure}[th!]
\begin{center}
\includegraphics[width=4.5in]{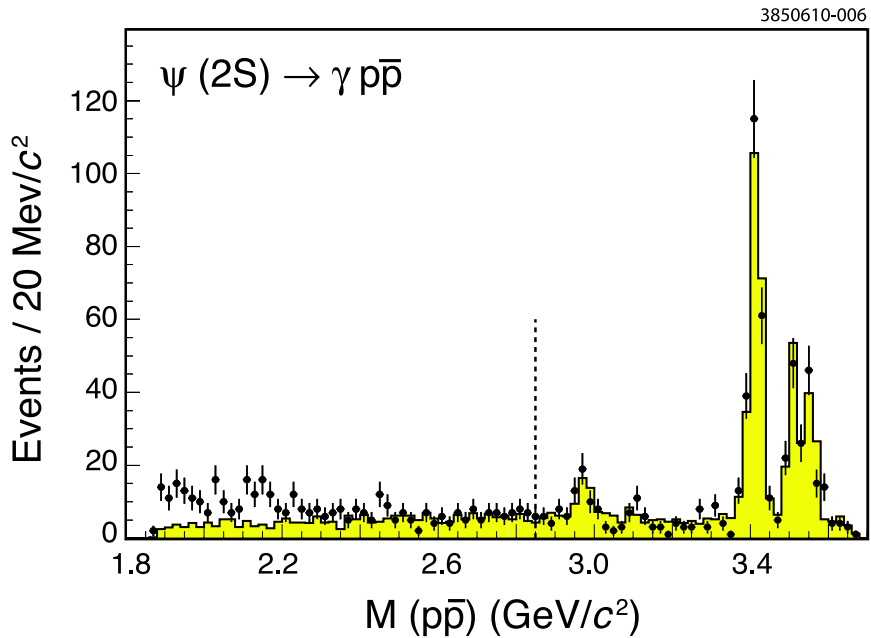}
\end{center}
\caption{
Distribution of $M(p\bar{p})$ for $\psi(2S)\to \gamma p\bar{p}$. 
All event selection criteria have been applied. 
The points represent data and the shaded histogram is the generic MC distribution, which
is normalized to the 24.5 million $\psi(2S)$ events in data.
The dashed line indicates the 
$M(p\bar{p})=2.85$ GeV/$c^{2}$  limit of the range used for 
$\psi(2S)\to \gamma p\bar{p}$ branching fraction calculations.
}
\end{figure}

\begin{figure}[th!]
\begin{center}
\includegraphics[width=4.5in]{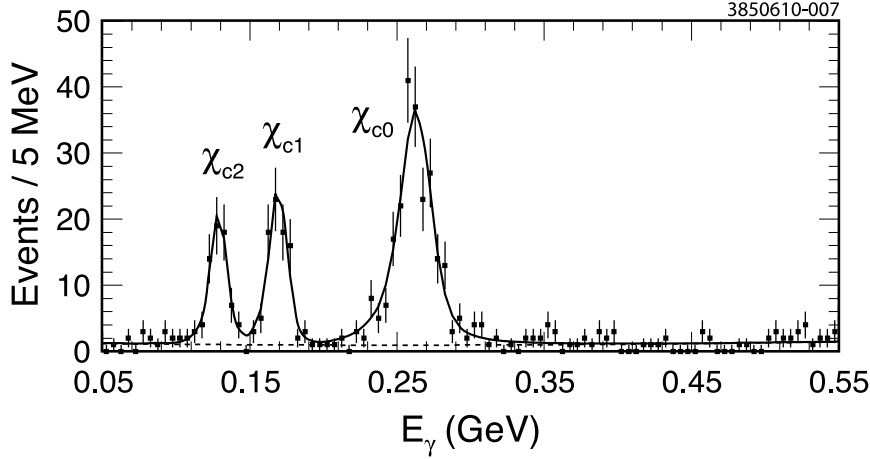}
\end{center}
\caption{
Fit of the photon energy distribution for $\psi(2S)\to \gamma p\bar{p}$.
Events in the figure are the same as in Fig.~6. The fit has $\chi^{2}$/d.o.f. = 100/98.
}
\end{figure}

In Fig.~6 we note that there is an excess of events in the data over the generic MC
simulation in the region $M(p\bar{p})<2.3$ GeV/$c^{2}$ . The generic MC simulation has 
no input for possible resonances in this mass region. 

The excitation of the $\chi_{cJ}$ states, shown in Fig.~6 gives us an 
opportunity to further test the appropriateness of our event selection.

To extract $\chi_{cJ}$ branching ratios we fit the photon energy $E_{\gamma}$ 
distribution shown in 
Fig.~7 with Breit--Wigner functions convolved with Crystal Ball line shape~\cite{crysball}, and a
second-order polynomial background. The fit results for photon energies agree
with those expected for the $\chi_{cJ}$ resonances within $\pm 2$ MeV. 
As seen in Table~II, our calculated values of $\mathcal{B}(\chi_{cJ} \to p\bar{p})$ 
agree within errors with both the results of a recent CLEO measurement~\cite{cleopp},
which use the same data, and PDG08~\cite{pdg08}. 
These measurements are intended as checks on the analysis
technique only and not as new measurements, and no systematic errors are included.

In order to explore the intermediate state resonances which are excited in
the reaction $\psi(2S)\to \gamma p\bar{p}$ we study several presentations
of the data. These include the $p \gamma$ versus $\bar{p} \gamma$ Dalitz plot,
the $M(p\bar{p})$ and $M(p\gamma)$ projections, and the 
distributions of $\cos \Theta$, where $\Theta$ is the angle between the proton 
and antiproton in the rest frame of the photon-proton system.

The three panels in Fig.~8 show $M^{2}(p \gamma)$ versus 
$M^{2}(\bar{p} \gamma)$ Dalitz plots respectively
for phase space MC simulation, data, and MC simulation with the intermediate resonances 
as described below. 
Figure~9 shows phase space MC distributions superimposed on the data 
for $M(p\bar{p})$, $M(p \gamma,\bar{p} \gamma)$ and $\cos \Theta$. (In the 
last two plots events have been included for both $p\gamma$ and $\bar{p}\gamma$,
resulting in double counting).

\begin{figure*}[tb!]
\begin{center}
\includegraphics[width=5.4in]{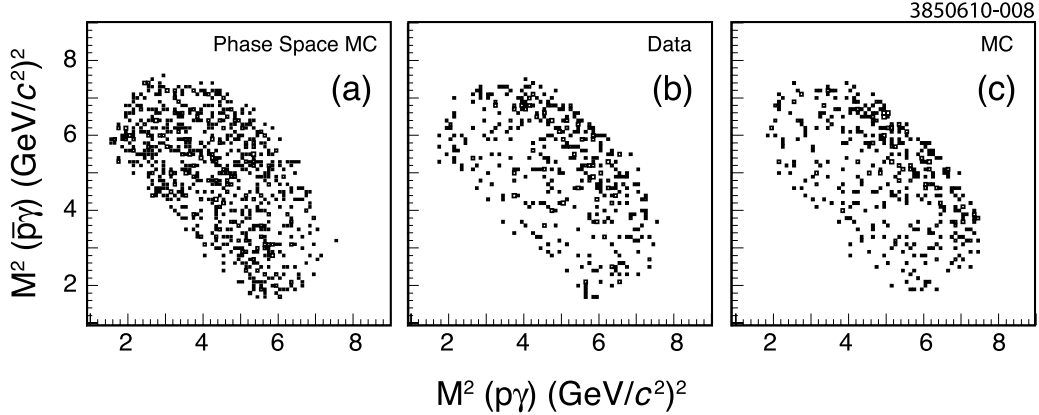}
\end{center}
\caption{Dalitz plots of $M^{2}(p\gamma)$ versus $M^{2}(\bar{p}\gamma)$
for $\psi(2S)\to \gamma p\bar{p}$:
(a) the phase space MC simulation; (b) the data; (c) the sum of three MC plots for
$f_{2}(1950)$, $f_{2}(2150)$ and phase space.}
\end{figure*}

\begin{figure*}[tb!]
\begin{center}
\includegraphics[width=5.4in]{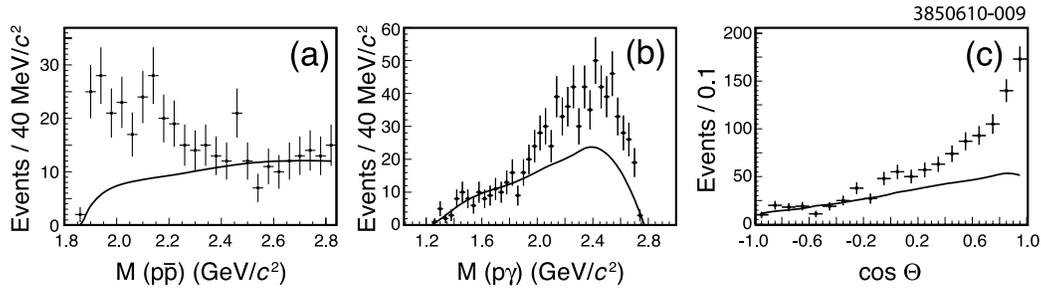}
\end{center}
\caption{Distributions in data (points) compared to the phase space
MC distributions (solid lines):
(a) the $M(p\bar{p})$ invariant mass distributions;
(b) the $M(p\gamma)$ distributions;
(c) the $\cos \Theta$ distributions. Phase space normalization is arbitrary in all plots.}
\end{figure*}

\begin{figure*}[!tb]
\begin{center}
\includegraphics[width=5.4in]{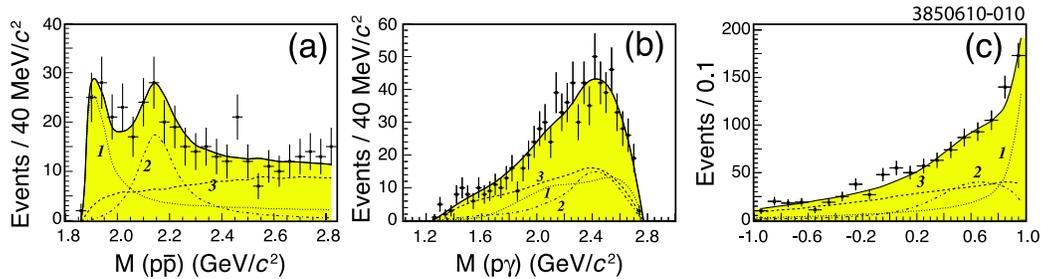}
\end{center}
\caption{Distributions in data (points) compared with the sum of the
MC distributions in the proportions given in Table~III (solid lines):
(a) the $M(p\bar{p})$ invariant mass distributions;
(b) the $M(p\gamma)$ distributions;
(c) the $\cos \Theta$ distributions.
The individual contributions are:
$f_{2}(1950)$ shown with the dotted line marked 1,
$f_{2}(2150)$ shown with the dotted-dashed line marked 2, and
phase space shown with the dashed line marked 3.}
\end{figure*}

In the distributions shown in Figs.~8 and 9, it is clear that pure phase space 
distributions fail
to describe the data. Significant contributions by intermediate states are required.
We have made MC studies of the contributions that various known scalar and
tensor meson resonances would make to these distributions. We find that the
best candidates are $f_{2}(1950)$ and $f_{2}(2150)$ with parameters given in PDG08. 
We determine MC shapes
of the contributions that $f_{2}(1950)$, $f_{2}(2150)$, and phase space make
to the distributions for $M(p\bar{p})$, $M(p \gamma)$ and $\cos \Theta$, and determine 
their relative magnitudes by fitting the data distributions. 
Using the PDG08~\cite{pdg08} values for masses and widths for the two resonances, the
best fit is obtained with relative fractions listed in Table~III. The corresponding MC
determined efficiencies, which are found to be insensitive to the uncertainties in
masses and widths of the resonances, are also listed in the table.
The overall efficiency of the admixture of the resonances and phase space is
\begin{eqnarray}
\langle \epsilon \rangle = 0.336\pm 0.008.
\end{eqnarray}

\begin{table}[!t]
\begin{center}
\caption{Fractions and efficiencies for the intermediate resonances and the phase space contribution
for the best fits for the reaction $\psi(2S)\to \gamma p\bar{p}$.}
\begin{tabular}{lcccc}
\hline \hline
State & M (MeV/$c^{2}$) & $\Gamma$ (MeV/$c^{2}$) & Fraction ($\%$) & 
$\epsilon$\\ \hline
$f_{2}(1950)$ & 1944$\pm$12 & 472$\pm$18 & 32$\pm$5 & 0.375 \\
$f_{2}(2150)$ & 2156$\pm$11 & 167$\pm$30 & 21$\pm$5 & 0.410 \\
Phase space & & & 47$\pm$6 & 0.277 \\
\hline \hline
\end{tabular}
\end{center}
\end{table}

As shown in Fig.~10 good fits to all three distributions are obtained
with the above admixtures, their respective 
$\chi^{2}$/d.o.f. being 15/24 for $M(p\bar{p})$, 33/35 for $M(p \gamma)$, and
23/20 for $\cos \Theta$. The resulting Dalitz plot, shown in Fig.~8(c), is also in good 
qualitative agreement with that for the data. No evidence is found in the data for
a narrow resonance $R$ with $\Gamma_{R} < 40$ MeV/$c^{2}$ anywhere in the region
$M(p\bar{p})=2200 - 2800$ MeV/$c^{2}$. The 90$\%$ confidence level upper limit is
$\mathcal{B}(\psi(2S)\to \gamma R)\times \mathcal{B}(R\to p\bar{p})<2\times 10^{-6}$.

\subsection{Determination of $\bm{\mathcal{B}(\psi(2S)\to \gamma p\bar{p})}$}

In the region $M(p\bar{p})<2.85$ GeV/$c^{2}$, we obtain $N=407\pm 20$ 
$\psi(2S)\to \gamma p\bar{p}$ candidate events. We evaluate
the background due to $\psi(2S)\to \pi^{0}p\bar{p}$, in which one photon from the
$\pi^{0}$ decay is lost, as $N_\mathrm{bkg}(\pi^{0})=38\pm 2$. In addition, by analyzing
the continuum data at $\sqrt{s}=3.76$ GeV we determine that the luminosity-normalized
continuum background contribution is $N_\mathrm{cont}=26\pm 8$ counts. 
With the relative contributions of $f_{2}(1950)$, $f_{2}(2150)$, and phase space as in
Table~III, and the effective overall efficiency (Eq.~(1)) we get
\begin{eqnarray}
\nonumber \mathcal{B}(\psi(2S)\to p \bar{p} \gamma) & = & \frac{N - N_\mathrm{bkg}-N_\mathrm{cont}}{\langle\epsilon \rangle \times N_{\psi(2S)}} \\ 
& = & (4.18\pm0.26\mathrm{(stat)})\times 10^{-5}.
\end{eqnarray}
The individual product branching fractions are
\begin{eqnarray}
& \nonumber \mathcal{B}(\psi(2S)\to \gamma f_{2}(1950))\times\mathcal{B}(f_{2}(1950)\to p\bar{p}) & \\ 
& = (1.2\pm0.2\mathrm{(stat)})\times 10^{-5}, &
\end{eqnarray}
\begin{eqnarray}
& \nonumber \mathcal{B}(\psi(2S)\to \gamma f_{2}(2150))\times\mathcal{B}(f_{2}
(2150)\to p\bar{p}) & \\
& = (0.72\pm0.18\mathrm{(stat)})\times 10^{-5}. &
\end{eqnarray}
Estimates of systematic errors are provided in Sec.~IX.

Our result for $\mathcal{B}(\psi(2S)\to \gamma p\bar{p})$ differs by 2$\sigma$ from 
the PDG08 result based on the BES measurement~\cite{besrad}, 
$\mathcal{B}(\psi(2S)\to \gamma p\bar{p})=(2.9\pm 0.6)\times 10^{-5}$ in which 
no account of intermediate resonances was taken. The results in Eqs.~(3) and (4) 
represent the first measurements of these product branching fractions.

\subsection{Search for Threshold Enhancement in $\bm{\psi(2S)\to \gamma p\bar{p}}$}

\begin{figure}[!ht] 
\begin{center} 
\includegraphics[width=4.5in]{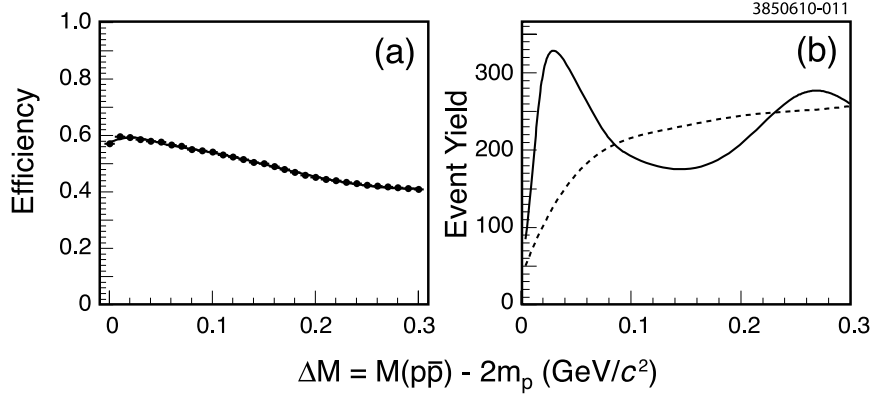} 
\end{center} 
\caption{ 
(a) MC-determined efficiency as a function of $\Delta M=M(p\bar{p})-2m_{p}$ for 
$\psi(2S) \to \gamma R_\mathrm{thr}$, $R_\mathrm{thr} \to p\bar{p}$.  
(b) The solid curve is the shape of the $\Delta M$ distribution from MC simulation 
for the admixtures shown in Table~III, and the dashed curve is the shape of the  
$\Delta M$ distribution from MC simulation for phase space alone.  
Relative normalizations are arbitrary.  
} 
\end{figure} 

\begin{figure}[!ht] 
\begin{center} 
\includegraphics[width=3.4in]{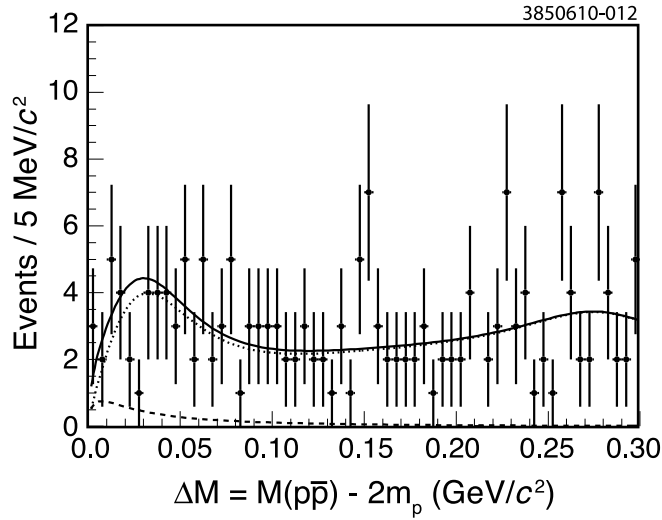} 
\end{center} 
\caption{ 
Fits of the $\Delta M \equiv M(p\bar{p})-2m_{p}$ distribution for 
$\psi(2S)\to \gamma p\bar{p}$ decay. The dotted line is the sum of the resonance 
and phase space contributions according to Table~III, the dashed line shows the fitted  
threshold resonance contribution. The solid line is the sum of all contributions.} 
\end{figure} 

Figure~10(a) shows that a good fit to the $M(p\bar{p})$ spectrum is obtained
with the sum of contributions from $f_{2}(1950)$, $f_{2}(2150)$, and phase space,
with $\chi^{2}$/d.o.f. = 15/24. No threshold resonance seems to be needed. However,
to reach a quantitative conclusion we study in detail the $p\bar{p}$ threshold
region, $\Delta M=M(p\bar{p})-2m_{p}=0-300$ MeV/$c^{2}$. To do so, we evaluate the 
contributions of the $f_{2}$ resonances and phase space in this region, and determine 
the efficiency for $\psi(2S)\to\gamma R_\mathrm{thr}$, $R_\mathrm{thr}\to p\bar{p}$ 
in this region. The results are shown in Fig.~11.

The $\Delta M=M(p\bar{p})-2m_{p}$ event distribution is shown in Fig.~12. A visual
inspection of the distributions shows that there is no evidence for a statistically
significant enhancement at the threshold, $\Delta M=0$. In fact, a straight line fit 
to the data gives $\chi^{2}$/d.o.f. = 52/58. However, we must consider the 
contributions
due to the $f_{2}$ resonances and phase space as has been determined in Table~III, and
as shown in Fig.~11(b),
and the efficiency $\langle \epsilon \rangle = 55.8\%$ in the threshold region. 
Figure~12 shows the best fit obtained 
using these contributions plus a Breit-Wigner threshold resonance with the parameters
obtained by BES~\cite{bes}, namely
$M(p\bar{p})=1859$ MeV/$c^{2}$, and $\Gamma =20$ MeV/$c^{2}$. 
The fit has $\chi^{2}$/d.o.f. = 53/58,
and includes the best fit threshold resonance $R_\mathrm{thr}$ with $9^{+10}_{-9}$ counts. 
This leads to
\begin{eqnarray} 
& \nonumber \mathcal{B}(\psi(2S)\to\gamma R_\mathrm{thr})\times\mathcal{B}(R_\mathrm{thr}\to p\bar{p})=
\frac{N_{R}}{\epsilon N_{\psi(2S)}} & \\ 
& = (0.66^{+0.73}_{-0.66})\times 10^{-6}~\mathrm{or}~<1.6\times 10^{-6},~90\%~\mathrm{CL}, & 
\end{eqnarray}
where CL means confidence level.
This is more than a factor three more restrictive than the current best 
limit~\cite{besrad}.

\section{The Decay $\bm{\psi(2S)\to \pi^{0} p\bar{p}}$}

Our analysis of $\psi(2S)\to \pi^{0}p\bar{p}$ follows the same steps as described
in Sec.~IV for $\psi(2S)\to \gamma p\bar{p}$. Figure~13 shows the three Dalitz plots
respectively for
(a) phase space MC simulation, (b) data, and (c) MC simulation with the resonances 
described below.
The phase space and data Dalitz plots differ dramatically, and the MC plot with the
resonances described below is in impressive agreement with the data. Figure.~14  
shows the projected distributions
for $M(p\pi^{0})$, $M(p\bar{p})$, and $\cos \Theta$, the polar angle of
$p$ in the rest frame of $\pi^{0}p$. It is clear that the pure phase space
distributions do not reproduce the data in either the Dalitz plots or the projected
distributions. Figure~15 shows the
same three distributions with good quality fits based on resonance shapes determined from
MC simulations, as described below. 

The $N^{*}$ intermediate states in $\psi(2S) \to \pi^{0}p\bar{p}$ are most clearly visible  
in the $M(p\pi^{0})$ distribution of Fig.~14(a), with enhancements near  
$M(p\pi^{0})\approx 1400$ MeV/$c^{2}$ and $M(p\pi^{0})\approx 2300$ MeV/$c^{2}$. 
Similarly, the meson intermediate states are most clearly visible in the $M(p\bar{p})$ 
distribution of Fig.~14(b), with enhancements near $M(p\bar{p})\approx 2100$ MeV/$c^{2}$ 
amd $M(p\bar{p})\approx 2900$ MeV/$c^{2}$. The enhancement at  
$M(p\pi^{0})\approx 1400$ MeV/$c^{2}$  can be identified with the well known 
$N^{*}(1440)$, which we call $N^{*}_{1}$, and the enhancement at 
$M(p\bar{p})\approx 2100$ MeV/$c^{2}$, which we 
call $R_{1}$, can be identified with the known resonance $f_{0}(2100)$~\cite{pdg08}. 
The large enhancements in $M(p\pi^{0})$ at 2300 MeV/$c^{2}$, which we call $N^{*}_{2}$, 
and in $M(p\bar{p})$ at 2900 MeV/$c^{2}$, which we call $R_{2}$, can not be identified 
with known $N^{*}$ and $f_{0,2}$ resonances, and we have to take an empirical approach 
for them. 
 
Because the mass and width of $f_{0}(2100)$ are well defined, in all subsequent analysis  
we keep them fixed to their PDG08 values. To determine the optimum values for the 
masses and widths of the $N^{*}_{1}$, $N^{*}_{2}$ and $R_{2}$ resonances following  
procedure was used. 

Because the Dalitz plot projections contain reflections, the projections 
can not be fitted with simple Breit-Wigner resonances. Instead, MC distributions have to 
be generated for individual resonances with assumed masses and widths, 
and their optimum values have to be determined by fitting the
data distributions with the MC generated distributions. Our procedure
takes account of reflections, but does not include taking account of any
possible interferences between resonances.

\begin{figure*}[tb!] 
\begin{center} 
\includegraphics[width=5.4in]{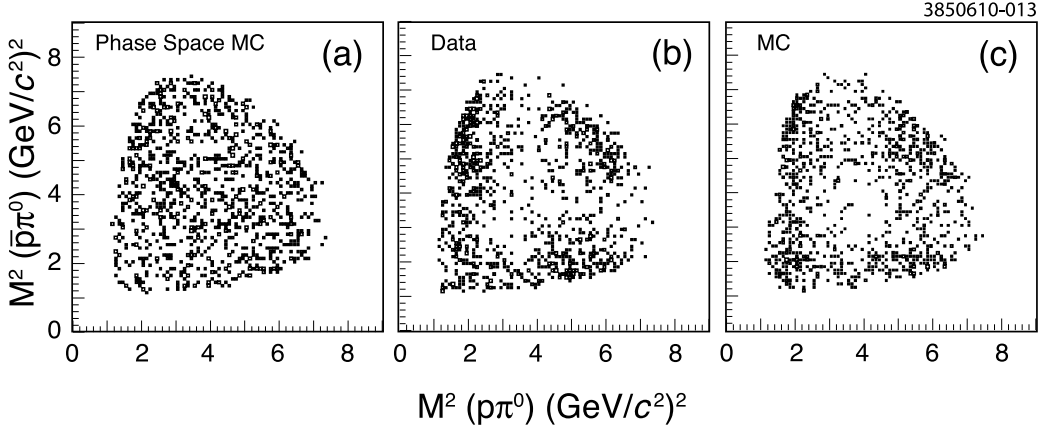} 
\end{center} 
\caption{Dalitz plots of $M^{2}(p\pi^{0})$ versus $M^{2}(\bar{p}\pi^{0})$ 
for the channel $\psi(2S)\to \pi^{0} p\bar{p}$: 
(a) the phase space MC simulation; (b) the data; (c) the sum of four MC plots for  
$R_{1}(2100)$, $R_{2}(2900)$, $N_{1}^{*}(1440)$ and $N_{2}^{*}(2300)$.} 
\end{figure*} 
 
\begin{figure*}[tb!] 
\begin{center} 
\includegraphics[width=5.4in]{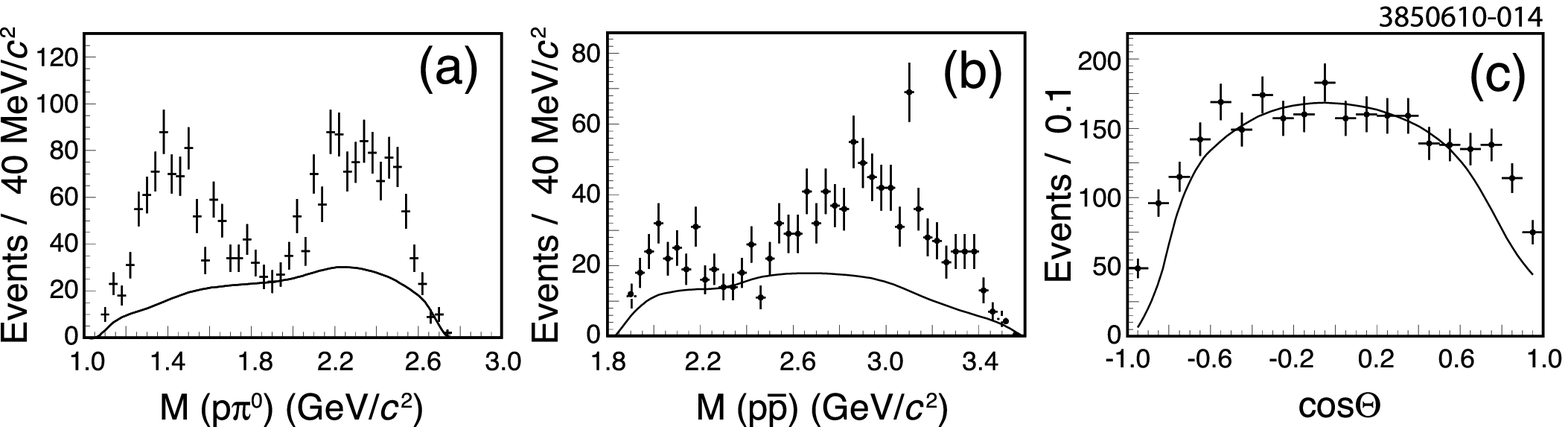} 
\end{center} 
\caption{Distributions in data (points) compared to the phase space 
MC distributions: 
(a) the $M(p\pi^{0})$ distributions. 
(b) the $M(p\bar{p})$ invariant mass distributions. 
(c) the $\cos \Theta$ distributions. 
Normalization is arbitrary in all plots.} 
\end{figure*} 
 
\begin{figure*}[tb!] 
\begin{center} 
\includegraphics[width=5.4in]{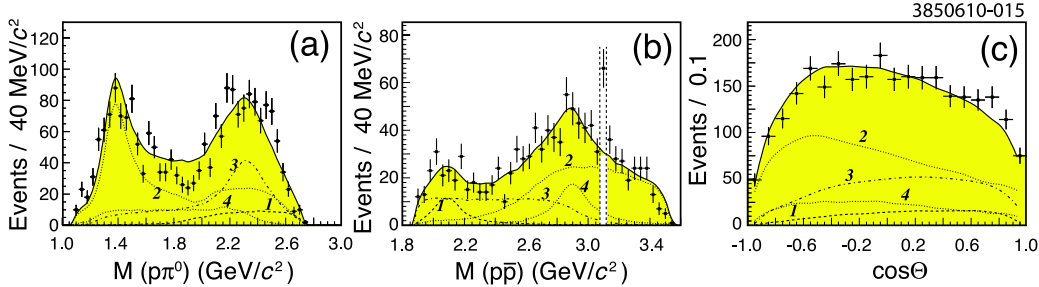} 
\end{center} 
\caption{Distributions in data (points) compared to the sum of the MC distributions 
in the proportions given in Table~IV (solid lines): 
(a) the $M(p\pi^{0})$ distributions; 
(b) the $M(p\bar{p})$ distributions; 
(c) the $\cos \Theta$ distributions. 
The individual contributions are: 
$R_{1}(2100)$ shown with the dashed line marked 1, 
$N_{1}^{*}(1440)$ shown with the dotted line marked 2, 
$N_{2}^{*}(2300)$ shown with the dotted-dashed line marked 3, 
$R_{2}(2900)$ shown with the dotted line marked 4.} 
\end{figure*} 

We first fit the $M(p\pi^{0})$ distribution with only $N^{*}_{1}$ and $N^{*}_{2}$ resonances 
and determine the best fit values for their masses and widths by iterating each in  
5 MeV/$c^{2}$ steps. We then fit the $M(p\bar{p})$ distribution with just the above  
$N^{*}_{1}$ and $N^{*}_{2}$ resonances. We find that the $M(p\bar{p})$ distribution is 
fitted poorly, with $\chi^{2}$/d.o.f.=65/37, and the enhancements at  
$M(p\bar{p})\approx 2100$ MeV/$c^{2}$ and $M(p\bar{p})\approx 2900$ MeV/$c^{2}$ are not 
reproduced. We then explicitly introduce fixed parameter $R_{1}(2100)$, and $R_{2}$ on 
whose parameters we iterate to find their best values. As expected, the fit to the 
$M(p\bar{p})$ distribution is improved, with $\chi^{2}$/d.o.f.=44/33. 
We go back to the $M(p\pi^{0})$ distribution to determine the effect of including  
$R_{1}$ and $R_{2}$. It is found that their contribution is structureless in the 
$M(p\pi^{0})$ distribution, and it does not 
affect the best fit parameters of $N^{*}_{1}$ and $N^{*}_{2}$.

In Table~IV final resonance parameters of $N^{*}_{1}$, $N^{*}_{2}$, $R_{1}$ and $R_{2}$ 
are listed. The errors in the masses and widths are those which change the likelihood of fits  
by two units. The efficiencies are as determined by MC simulations. The relative fractions 
are determined by the final fit to the $M(p\bar{p})$ distribution. 
 
\begin{table}[tb!]
\begin{center}
\caption{Fractions and efficiencies for the intermediate resonances for the 
reaction $\psi(2S)\to \pi^{0} p\bar{p}$.}
\begin{tabular}{lcccc}
\hline \hline
Resonance & M (MeV/$c^{2}$) & $\Gamma$ (MeV/$c^{2}$) & Fraction ($\%$) & $\epsilon$\\ \hline
$N_{1}^{*}(1440)$ & 1400$\pm$25 & 220$\pm$20 & 50$\pm$4 & 0.241 \\
$N_{2}^{*}(2300)$ & 2300$\pm$25 & 300$\pm$30 & 28$\pm$4 & 0.276 \\
$R_{1}(2100)$ & 2103$\pm$8 & 209$\pm$19 & 8$\pm$3 & 0.275 \\
$R_{2}(2900)$ & 2900$\pm$20 & 250$\pm$25 & 24$\pm$4 & 0.241 \\
\hline \hline
\end{tabular}
\end{center}
\end{table}

The fits obtained for $M(p\pi^{0})$, $M(p\bar{p})$ and $\cos \Theta$ distributions with  
the final set of parameters for all four resonances are shown in Fig.~15(a,b,c). The   
individual resonance contributions are shown with dotted and dashed lines.  
The corresponding composite Dalitz plot is shown in Fig.~13(c). It agrees very well with 
that for data in Fig.~13(b). No evidence
for a $p\bar{p}$ threshold enhancement is observed in the $M(p\bar{p})$ distribution
of Fig.~15(b).

BES reported the result for $\psi(2S)\to \pi^{0}p\bar{p}$ using their sample of 14 
million $\psi(2S)$ events~\cite{bespsip}. Since the number of events was almost a factor
of two smaller than in the present investigation, they were not able to reach 
any conclusions about intermediate states other than to note that there was 
``indication of some enhancement around 2 GeV/$c^{2}$.''

\subsection{Determination of $\bm{\mathcal{B}(\psi(2S)\to \pi^{0}p\bar{p})}$}

We consider all events in the $M(p\bar{p})$ spectrum for $M(p\bar{p})<$ 3.6 GeV/$c^{2}$ 
(Fig.~15(b)) for determination of the branching fraction 
$\mathcal{B}(\psi(2S)\to \pi^{0} p\bar{p})$, except those with 
$M(p\bar{p})=3.097\pm 0.020$ GeV/$c^{2}$, which could come from $J/\psi$ production.

We obtain $N=1063\pm33$ $\psi(2S)\to \pi^{0}p\bar{p}$ candidate events.
Using our branching fraction of 
$\mathcal{B}(\psi(2S)\to \gamma p\bar{p} = (4.18 \pm 0.26)\times 10^{-5}$, and the 
efficiency determined in Eq.~(1) we estimate $N_{\mathrm{bkg}}=15\pm 1$ background 
counts due to misidentified $\gamma p\bar{p}$ events.

We estimate the non-resonant contribution of $\pi^{0} p\bar{p}$ production
by using the data taken at the off--$\psi(2S)$ resonance energy of $\sqrt{s} = 3.67$ GeV. 
It leads to a luminosity-normalized non-resonant contribution
in our data, $N_\mathrm{cont}=105\pm 16$.

The efficiencies determined from the $N^{*}_{1}(1440)$, $N^{*}_{2}(2300)$, $R_{1}(2100)$
and $R_{2}(2900)$ MC simulations are 24.1$\%$, 27.6$\%$, 27.5$\%$, and 24.1$\%$, 
respectively.
The overall efficiency of the admixture of the resonances is $\langle \epsilon \rangle = (25.4\pm 0.2)\%$.

This yields a branching fraction of
\begin{eqnarray}
\nonumber \mathcal{B}(\psi(2S)\to p \bar{p} \pi^{0}) & = & \frac{N - N_{\mathrm{bkg}} - 
N_\mathrm{cont}} {\langle \epsilon \rangle \times N_{\psi(2S)} \times \mathcal{B}(\pi^{0} \to \gamma \gamma)}\\
& = & (1.54\pm0.06\mathrm{(stat)})\times 10^{-4}.
\end{eqnarray}
This result is in agreement with the PDG08~\cite{pdg08} value of 
$(1.33\pm 0.17)\times 10^{-4}$, and has a factor three smaller error.

We can also determine the product branching fractions for the 
$N^{*}_{1}(1440)$, $N^{*}_{2}(2300)$, $R_{1}(2100)$ and $R_{2}(2900)$ resonances by taking
account of their respective fractions and efficiencies given in Table~IV. 
The resulting product branching fractions are
\begin{eqnarray}
&\nonumber \mathcal{B}(\psi(2S)\to \bar{p}N^{*}_{1}(1440))\times\mathcal{B}(N^{*}_{1}
(1440) \to p\pi^{0}) & \\
& = (8.1\pm0.7\mathrm{(stat)})\times 10^{-5}, & 
\end{eqnarray}
\begin{eqnarray}
&\nonumber \mathcal{B}(\psi(2S)\to \bar{p}N^{*}_{2}(2320))\times\mathcal{B}(N^{*}_{2}
(2320)\to p\pi^{0}) & \\
& = (4.0\pm0.6\mathrm{(stat)})\times 10^{-5}, &
\end{eqnarray}
\begin{eqnarray}
&\nonumber \mathcal{B}(\psi(2S)\to \pi^{0}R_{1}(2100))\times\mathcal{B}(R_{1}(2100)\to 
p\bar{p}) & \\
& = (1.1\pm0.4\mathrm{(stat)})\times 10^{-5}, &
\end{eqnarray}
\begin{eqnarray}
&\nonumber \mathcal{B}(\psi(2S)\to \pi^{0}R_{2}(2900))\times\mathcal{B}(R_{2}(2900)\to 
p\bar{p}) & \\
& = (2.3\pm0.7\mathrm{(stat)})\times 10^{-5}. &
\end{eqnarray}

These are the first determinations of these product branching fractions.
Estimates of systematic errors are provided in Sec.~IX.

\section{The Decay $\bm{\psi(2S)\to \eta p\bar{p}}$}

As shown in Fig.~1, the yield of $\eta p\bar{p}$ (184 counts) is nearly a factor of 6 
smaller than for $\pi^{0} p\bar{p}$ (1063 counts). 
Dalitz plots of $M^{2}(p \eta)$ versus $M^{2}(\bar{p} \eta)$ are shown in Fig.~16, (a) for 
pure phase space, (b) for the data, and (c) for resonances described below.
As in the $\psi(2S)\to \pi^{0} p\bar{p}$ case, it is seen that the Dalitz plot for the 
data (Fig.~16(b)) is completely different from the uniformly populated Dalitz plot for 
phase space MC simulation (Fig.~16(a)). The data are clearly dominated by the contribution 
of intermediate states. As shown in Fig.~16(c) the  sum of MC simulated contributions 
of resonances described below reproduces the data very well.

\begin{figure*}[tb!] 
\begin{center} 
\includegraphics[width=5.6in]{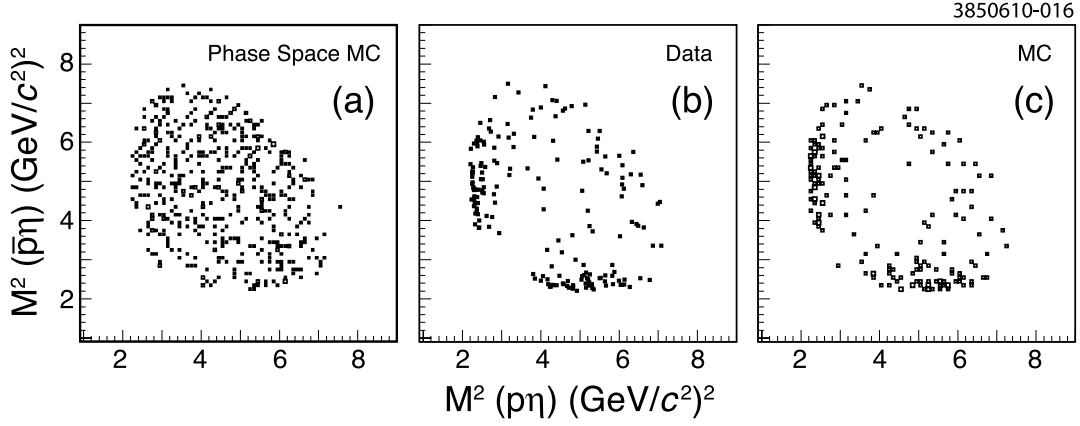} 
\end{center} 
\caption{Dalitz plots of $M^{2}(p\eta)$ versus $M^{2}(\bar{p}\eta)$ 
for the channel $\psi(2S)\to \eta p\bar{p}$: 
(a) the phase space MC simulation; (b) the data; (c) the sum of two MC plots for $R_{1}(2100)$ 
(20$\%$), and $N^{*}(1535)$ (80$\%$).} 
\end{figure*} 
 
\begin{figure*}[tb!] 
\begin{center} 
\includegraphics[width=5.7in]{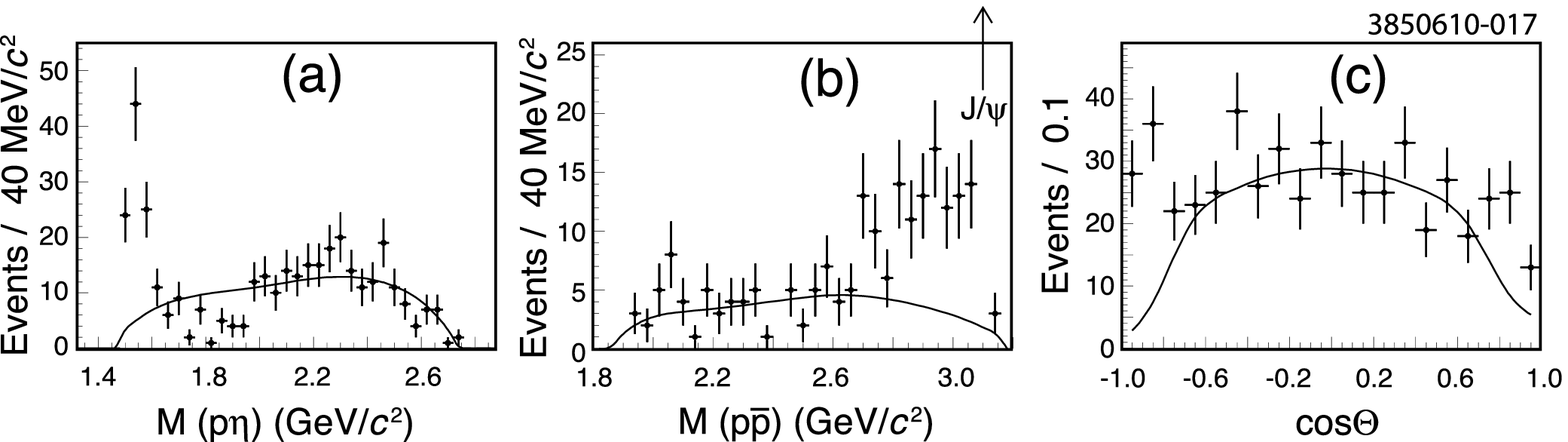} 
\end{center} 
\caption{Distributions in data (points) compared to the phase space 
MC distributions: 
(a) the $M(p\eta)$ distributions; 
(b) the $M(p\bar{p})$ invariant mass distributions; 
(c) the $\cos \Theta$ distributions. 
Normalization of the curves is arbitrary in all plots.} 
\end{figure*} 
 
\begin{figure*}[!tb] 
\begin{center} 
\includegraphics[width=5.7in]{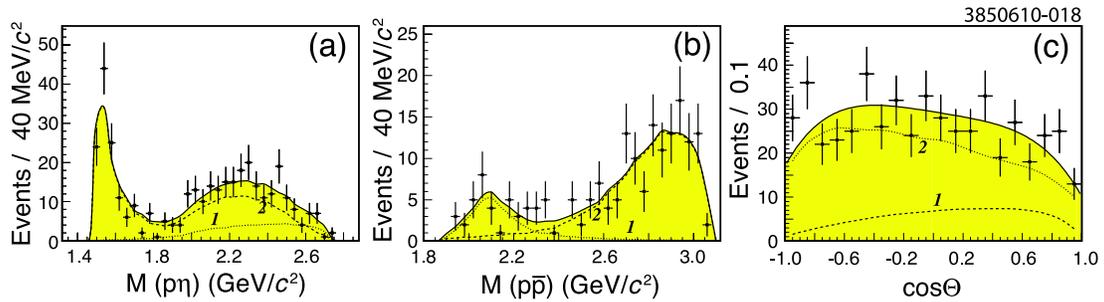} 
\end{center} 
\caption{Distributions in data (points) compared to the sum of the 
MC distributions in the proportions given in Table~V (solid lines): 
(a) the $M(p\eta)$ distributions; 
(b) the $M(p\bar{p})$ distributions; 
(c) the $\cos \Theta$ distributions. 
The individual contributions are: 
$R_{1}(2100)$ shown with the dotted line marked 1, 
$N^{*}(1535)$ shown with the dashed line marked 2.} 
\end{figure*} 

The projected distributions for (a) $M(p\eta)$, (b) $M(p\bar{p})$, and
(c) $\cos \Theta$, where $\Theta$ is the polar angle of the $p$ in the rest frame
of $\eta p$, are shown in Fig.~17, together with MC generated distributions
for phase space. As expected, the phase space distributions do not reproduce the data 
distributions.

In the $M(p\eta)$ invariant mass distribution (Fig.~17(a)) no evidence is found for 
$N^{*}_{1}(1440)$ and $N^{*}_{2}(2300)$ resonances seen in the $M(p\pi^{0})$ 
plot of Fig.~15(a), but a large peak is observed at $M(p\eta)\sim$ 1540 MeV/$c^{2}$. 
This suggests excitation of $N^{*}(1535)$ $J^{P}={1 \over 2}^{-}$ nucleon resonance 
of PDG08~\cite{pdg08} with $M = 1525-1545$ (MeV/$c^{2}$) and $\Gamma = 125-175$ (MeV/$c^{2}$), which is 
known to decay into $N\eta$ with a branching fraction of (45-60)$\%$.

In the $M(p\bar{p})$ distribution (Fig.~17(b)) there is a broad enhancement in the $2.7 - 3.0$ GeV/$c^{2}$ 
region which arises mainly as reflection of the $N^{*}(1535)$ resonance in
$M(p\eta)$. In addition there is a narrow enhancement near 
$M(p\bar{p})\approx 2100$ MeV/$c^{2}$  reminiscent of the one observed in $M(p\bar{p})$ from
$\psi(2S)\to \pi^{0}p\bar{p}$ decay.

The optimized masses and widths we obtain for these resonances, and their fractions and 
estimated efficiencies are shown in Table~V. It is found that the MC-determined 
efficiencies are insensitive to the uncertainties in masses and widths of the resonances. 

\begin{table}[!htb]
\begin{center}
\caption{Fractions and efficiencies for the intermediate resonances for the best fits for
the reaction $\psi(2S)\to \eta p\bar{p}$.}
\begin{tabular}{lcccc}
\hline \hline
Resonance & M (MeV/$c^{2}$) & $\Gamma$ (MeV/$c^{2}$) & Fraction ($\%$) & $\epsilon$\\ \hline
$N^{*}(1535)$ & 1535$\pm$10 & 150$\pm$25 & 80$\pm$6 & 0.294 \\
$R_{1}(2100)$ & 2103$\pm$8 & 209$\pm$19 & 20$\pm$6 & 0.259 \\
\hline \hline
\end{tabular}
\end{center}
\end{table}

We fit our data distribution for $M(p\eta)$ and $M(p\bar{p})$ with an admixture of MC 
simulated shapes for these two resonances (the fit result shows that the 
contribution from the phase space MC is consistent with zero). 
The best fit admixture is: 
\begin{equation} 
(0.20\pm 0.06)\times R_{1}(2100) + (0.80\pm 0.06)\times N^{*}(1535). 
\end{equation} 

The final results for the $M(p \eta)$, $M(p\bar{p})$ and $\cos \Theta$
distributions are presented in Fig.~18 with solid lines.
The dashed and dotted lines represent the contributions from the individual resonances.
Good agreement between the data and the fitted distributions is obtained for all three
distributions, with $\chi^{2}$/d.o.f. are 35/30 $(M(p \eta))$, 30/30 $(M(p\bar{p}))$
and 32/20 ($\cos \Theta$).

The $M^{2}(p\eta)$ versus $M^{2}(\bar{p}\eta)$ Dalitz plot in Fig.~16(c) constructed with
the MC distributions for the above resonance admixture is seen to reproduce very well 
the Dalitz plot of data in Fig.~16(b). 

\subsection{Determination of $\bm{\mathcal{B}(\psi(2S)\to \eta p\bar{p})}$}

We consider the entire
$M(p\bar{p})$ spectrum with $M(p\bar{p})< 3.077$ GeV/$c^{2}$  in Fig.~18(b) for the 
determination of $\mathcal{B}(\psi(2S)\to \eta p\bar{p})$.

We obtain $N=184\pm14$ $\psi(2S)\to \eta p\bar{p}$ candidate events.
We do not find any background contribution from feed-down from other decay channels.
We estimate the continuum contribution of $\eta p\bar{p}$ production by using the data
taken at $\sqrt{s} = 3.67$ GeV. It leads to the luminosity-normalized contribution
$N_\mathrm{cont} = 30\pm8$ counts.

The reconstruction efficiencies determined from the $R_{1}(2100)$ and $N^{*}(1535)$ 
MC simulations are 25.9$\%$ and
29.4$\%$, respectively and the overall effective efficiency of the resonances admixture is
$\langle \epsilon \rangle = (28.7\pm 0.2)\%$.

This yields a branching fraction of
\begin{eqnarray}
\nonumber \mathcal{B}(\psi(2S)\to \eta p\bar{p}) & = & \frac{N - N_{\mathrm{cont}}}
{\langle \epsilon \rangle \times N_{\psi(2S)} \times \mathcal{B}(\eta \to \gamma \gamma)}\\
& = & (5.6\pm0.6\mathrm{(stat)})\times 10^{-5}.
\end{eqnarray}
This is in agreement with the PDG08~\cite{pdg08} value of $(6.0\pm 1.2)\times 10^{-5}$, and
has a factor two smaller error.
We can determine the product branching ratios for the production of $N^{*}(1535)$
and $R_{1}(2100)$ resonances by taking account of their respective fractions and efficiencies.
We obtain
\begin{eqnarray}
&\nonumber \mathcal{B}(\psi(2S)\to \bar{p}N^{*}(1535))\times\mathcal{B}(N^{*}(1535)\to
p\eta) & \\
& =(4.4\pm 0.6\mathrm{(stat)})\times 10^{-5}, &
\end{eqnarray}
\begin{eqnarray}
\nonumber \mathcal{B}(\psi(2S) & \to & \eta R_{1}(2100)\times\mathcal{B}(R_{1}(2100)\to
p\bar{p}) \\
& = & (1.2\pm 0.4\mathrm{(stat)})\times 10^{-5}. 
\end{eqnarray}

These are the first determinations of these product branching fractions.
Estimates of systematic errors are provided in Sec.~IX.

\section{Search for $\bm{p\bar{p}}$ Threshold Enhancement in $\bm{J/\psi \to \gamma p\bar{p}}$}

Although the number of $\pi^{+}\pi^{-}$ tagged $J/\psi$ events in our sample of 
$\psi(2S)\to \pi^{+}\pi^{-}J/\psi$ is 8.7 million, as compared
to the 58 million event $J/\psi$ sample of BES II, it is instructive to analyze it for
the sub-threshold resonance with $M(p\bar{p})=1859^{+3}_{-10}$$^{+5}_{-25}$ MeV/$c^{2}$
reported by BES~\cite{bes}.

For selection of $\psi(2S) \to \pi^{+}\pi^{-}J/\psi$,
$J/\psi \to \gamma p\bar{p}$ events, we require first $\chi^{2}_{\mathrm{p\bar{p}~vertex}}<20$ and
$\chi^{2}_{\mathrm{fit}}/\mathrm{d.o.f.}<3$ of the $J/\psi$ mass constrained fit to $p\bar{p}$ and most
energetic shower in the event, then $\chi^{2}_\mathrm{vertex}<40$ and
$\chi^{2}_\mathrm{fit}/\mathrm{d.o.f.}<10$ of the four-momentum conservation constrained fit to
$\pi^{+}\pi^{-}$ and $J/\psi$.

In Fig.~19 we show the distribution of $p\bar{p}$ invariant
mass as a function of $\Delta M=M(p\bar{p})-2m_{p}$ in the extended mass region
$\Delta M=$ 0 - 970 MeV/$c^{2}$. We believe that it is essential to analyze the data in
the extended mass region, because as we have seen for $\psi(2S)$ decays, higher mass
resonances make contributions all the way down to the $p\bar{p}$ threshold. 
Further, as shown in Fig.~19, a much better estimate of the phase space contribution can 
be made when data in the extended mass region is taken into account.
Fig.~19 shows an enhancement near $p\bar{p}$ threshold and
a large broad enhancement 
around $\Delta M \approx 200$ MeV/$c^{2}$.
We therefore analyze our data in the extended mass
region, $\Delta M = 0-970$ MeV/$c^{2}$, and take account of possible resonances other than the
one near the $p\bar{p}$ threshold. Our analysis differs in this essential respect from
that of BES in which data in the limited region, $\Delta M=0-300$ MeV/$c^{2}$, was
analyzed, and no account was taken of the enhancement around 
$\Delta M \approx 200$ MeV/$c^{2}$.

\begin{figure}[!tb]
\begin{center}
\includegraphics[width=3.2in]{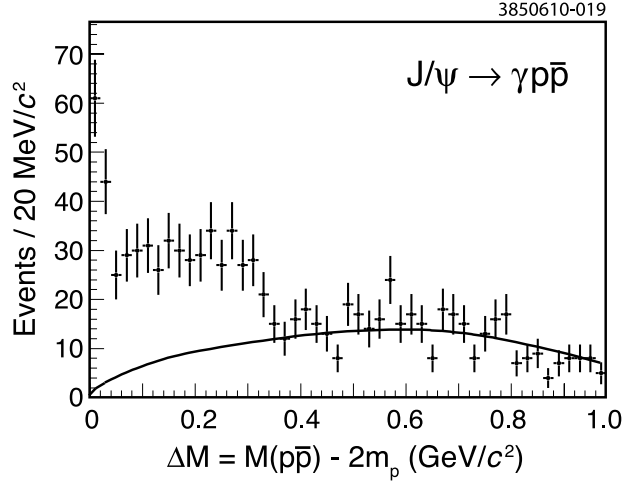}
\end{center}
\caption{
The $\Delta M=M(p\bar{p})-2m_{p}$ invariant mass distribution for the data
for $J/\psi \to \gamma p\bar{p}$. The curve illustrates the shape of the phase space
contribution.
}
\end{figure}

\begin{figure}[!tb]
\begin{center}
\includegraphics[width=3.2in]{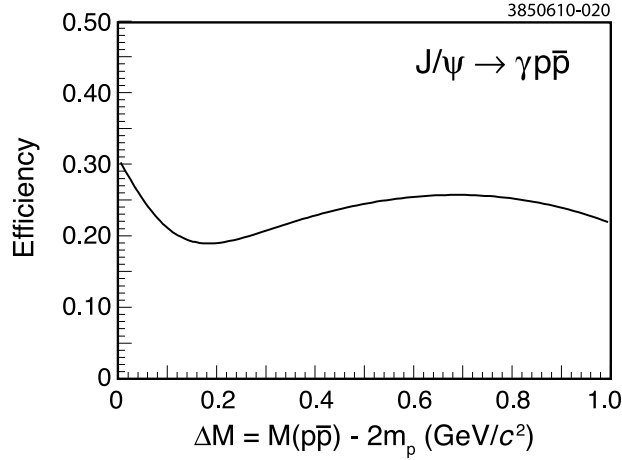}
\end{center}
\caption{
MC-determined efficiency as a function of $\Delta M \equiv M(p\bar{p})-2m_{p}$
for $J/\psi \to p\bar{p}$ decays. The weighted average efficiency over the whole
range is 25.4$\%$.
}
\end{figure}

We have made an attempt to fit the present $\Delta M$ distribution with 
a threshold resonance plus the complement of resonances and phase space observed
in the case of $\psi(2S)\to \gamma p\bar{p}$, {\it i.e.} $f_{2}(1950)$ (corresponding to
$\Delta M \approx 74$ MeV/$c^{2}$), $f_{2}(2150)$ (corresponding to $\Delta M \approx 224$ MeV/$c^{2}$), and
phase space. No evidence for a contribution due to the $f_{2}(1950)$ resonance was found. 
All subsequent fits were therefore tried with an S-wave threshold resonance plus MC shapes 
determined for contributions of a resonance at $M = 2100\pm 20$ (MeV/$c^{2}$), 
$\Gamma = 160\pm 20$ (MeV/$c^{2}$) (our optimum values), and phase space. 
The MC-determined event 
selection efficiency as a function of $\Delta M$ is shown in Fig.~20.
The average efficiency, weighted by the threshold resonance contribution, as shown by the
curves marked (1) in Fig.~21, was found to be $\epsilon = 0.254$.

\begin{figure}[!tb]
\begin{center}
\includegraphics[width=3.2in]{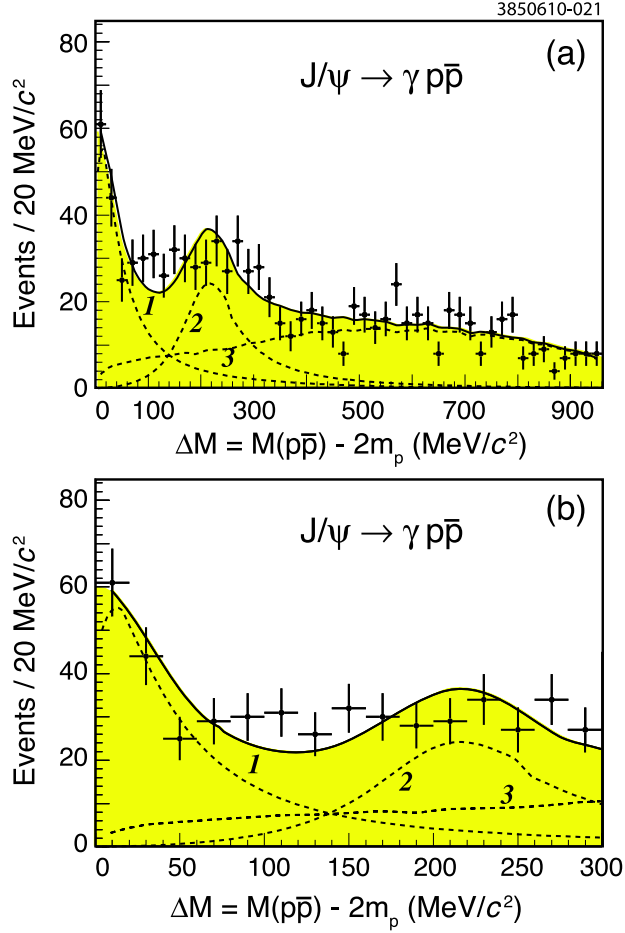}
\end{center}
\caption{
Fits of the $\Delta M=M(p\bar{p})-2m_{p}$ invariant mass distribution for
$J/\psi \to \gamma p\bar{p}$ decays. Dashed lines are
the contributions of (1) $R_\mathrm{thr}$, (2) $f_{2}(2100)$ and (3) the phase space.
The solid line is the sum of all three contributions. (a) Fit in the full region
$\Delta M=0-970$ MeV/$c^{2}$. (b) Same fit in the $\Delta M = 0-300$ MeV/$c^{2}$.
}
\end{figure}

We find that if, like BES, we fit the $\Delta M$ distribution only in the region
$\Delta M = 0 - 300$ MeV/$c^{2}$, and do not consider the contribution due to the resonance
at $M(p\bar{p})=2100$ MeV/$c^{2}$, we obtain a good fit ($\chi^{2}/\mathrm{d.o.f.}$=15/26)
which is essentially identical
to that obtained by BES~\cite{bes}, with the results 
\begin{align}
\nonumber & M(R_\mathrm{thr}) = 1861^{+16}_{-6}~ \mathrm{MeV}/c^{2}, \\
\nonumber & \Gamma(R_\mathrm{thr}) = 0^{+32}_{-0}~ \mathrm{MeV}/c^{2},  \\
& \mathcal{B}(J/\psi \to \gamma R_{\mathrm{thr}})\!\times\! \mathcal{B}(R_{\mathrm{thr}}\to p\bar{p}) = (5.9^{+2.8}_{-3.2})\!\times\! 10^{-5}.
\end{align}
The errors are statistical only.

Fitting data in the extended region $\Delta M = 0 - 970$ MeV/$c^{2}$ 
leads to a better determination of the phase space contribution; it is smaller 
than what is obtained if the fit is confined to the small region, 
$\Delta M = 0 - 300$ MeV/$c^{2}$.
The fit with all parameters kept free ($\chi^{2}/\mathrm{d.o.f.}$=97/89)
leads to the fractional contributions of $f_2(2100)$ of $(22\pm3)\%$ and 
phase~space$\,(54\pm3)\%$.  The threshold resonance obtains $231^{+85}_{-59}$ counts 
and leads to the following parameters
\begin{align}
\nonumber & M(R_\mathrm{thr}) = 1837^{+10+9}_{-12-7}~\mathrm{MeV}/c^{2}, \\
\nonumber & \Gamma(R_\mathrm{thr}) = 0^{+44}_{-0}~ \mathrm{MeV}/c^{2}, \\
& \mathcal{B}(J/\psi \to \gamma R_\mathrm{thr})\!\times\! \mathcal{B}(R_\mathrm{thr}\to p\bar{p}) = (11.4^{+4.3+4.2}_{-3.0-2.6})\!\times\! 10^{-5}. 
\end{align}
The first errors are statistical, and the second errors are estimates of systematic
errors obtained by varying the mass and width of the 2100 MeV/$c^{2}$ resonance by
their 1$\sigma$ uncertainties.
The same result, $M(R_\mathrm{thr})=1837^{+9}_{-12}$ MeV/c$^{2}$, is obtained when 
$\Gamma(R_\mathrm{thr})$ is fixed to 20 MeV/$c^{2}$.

Figure~21(a) shows the fit in the region $\Delta M=0-970$ MeV/$c^{2}$, and Fig.~21(b)
shows the same fit in the $\Delta M=0-300$ MeV/$c^{2}$ region.

The fit result for mass in Eq.~16 is consistent with the conjecture that the threshold 
enhancement might be due to the tail of a subthreshold resonance at that mass. 
This possibility was raised earlier by BES with their observation of a resonance with mass 
$M(R)=1833.7\pm 6.1\pm2.7$ MeV/$c^{2}$ in the reaction $J/\psi \to \gamma R$, $R\to 
\pi^{+}\pi^{-}\eta$~\cite{bes1837}.

\section{Systematic Uncertainties} 
 
In Table~VI we list various contributions to the systematic uncertainties in the branching 
fractions. References to previous CLEO studies for several of these are also given. 
The uncertainty in feed-down and continuum contributions to the background leads to systematic 
uncertainties of 1$\%$, 1.4$\%$, and 2$\%$ in $\gamma p\bar{p}$, $\pi^{0} p\bar{p}$, and  
$\eta p\bar{p}$, respectively. All the above contributions add in quadrature to 3.5$\%$, 
4.0$\%$, and 5.5$\%$ uncertainty in $\gamma p\bar{p}$, $\pi^{0} p\bar{p}$, and  
$\eta p\bar{p}$, respectively. Since the uncertainties in the fractions of individual  
resonance and continuum contributions are taken into account in the statistical  
uncertainties, the above are the systematic uncertainties in the product branching  
fractions for the individual resonances in these decays. For  
$\mathcal{B}(\psi(2S)\to \gamma p\bar{p})$, $\mathcal{B}(\psi(2S)\to \pi^{0} p\bar{p})$,  
and $\mathcal{B}(\psi(2S)\to \eta p\bar{p})$, the uncertainties in the fractions of  
individual resonance and continuum contributions lead to additional systematic uncertainties 
 because the different contributions have different efficiencies. To take account of  
correlations in the fractions, the effective overall efficiencies were determined by MC  
simulations and the relative uncertainties found to be 2.4$\%$, 0.8$\%$, and 0.7$\%$ for  
$\gamma p\bar{p}$, $\pi^{0} p\bar{p}$, and $\eta p\bar{p}$, respectively. Thus the total  
systematic uncertainties in $\mathcal{B}(\psi(2S)\to \gamma p\bar{p})$,  
$\mathcal{B}(\psi(2S)\to \pi^{0} p\bar{p})$, and $\mathcal{B}(\psi(2S)\to \eta p\bar{p})$  
are 4.2$\%$, 4.1$\%$, and 5.5$\%$, respectively.  
 
The results for the branching fractions with systematic errors are given in Table~VII. 

It is found that $\pm 100$ MeV/$c^{2}$ changes in the masses and widths of resonances
introduce changes in branching fractions much less than 1$\%$.

\begin{table}[th]
\begin{center}
\caption{Sources of systematic uncertainties in branching fractions in $\%$. References to 
previous CLEO publications for estimates of uncertainties are given in square brackets.}
\begin{tabular}{lccc}

\hline \hline
Source \hfill [Ref] & $\gamma p\bar{p}$ & $\pi^{0} p\bar{p}$ & $\eta p\bar{p}$ \\ \hline
Number of $\psi(2S)$\hfill \cite{cleopsip} & 2.0 & 2.0 & 2.0 \\
Trigger efficiency \hfill \cite{cleotrig,cleogam} & 1.0 & 1.0 & 1.0 \\
Tracking efficiency \hfill \cite{cleotrack} & 2$\times$1.0 & 2$\times$1.0 & 2$\times$1.0 \\
Particle identification \hfill \cite{cleopsip} & 1.0 & 1.0 & 1.0 \\
$\gamma$, $\pi^{0}$, $\eta$ reconstruction~\cite{cleogam} & 1.0 & 2.0 & 4.0 \\
Background subtraction & 1.0 & 1.4 & 2.0 \\
& & & \\
Sub total (quadrature) & 3.5 & 4.0 & 5.5 \\
& & & \\
Resonance fractions & 2.4 & 0.8 & 0.7 \\ \hline
Total (quadrature) & 4.2 & 4.1 & 5.5 \\ \hline \hline
\end{tabular}
\end{center}
\end{table}

\section{Summary and Conclusions}

Using CLEO data for 24.5 million $\psi(2S)$ we have studied decays
$\psi(2S)\to \gamma p\bar{p}$, $\pi^{0}p\bar{p}$, and $\eta p\bar{p}$. In all three
decays we find that intermediate $N^{*}(\bar{N}^{*})$ states decaying into
$\pi^{0}N(\bar{N})$ and $\eta N(\bar{N})$, and $f_{J}$, $a_{J}$ meson resonances decaying
into $p\bar{p}$ make important contributions to the total decay. We have determined
branching fractions for the total decay and for the contributions of the individual
intermediate states. For the total decays our branching fractions have factors two to
three smaller uncertainties than in the current literature. The product branching
fractions for decays through individual intermediate states have been determined for
the first time. The results are summarized in Table~VII.

We do not find any evidence for a threshold enhancement in any of the three $\psi(2S)$ 
decay channels. For $\psi(2S)\to \gamma p\bar{p}$ we set a stringent upper limit for the
threshold resonance $R_\mathrm{thr}$,
$\mathcal{B}(\psi(2S)\to\gamma R_\mathrm{thr})\times\mathcal{B}(R_\mathrm{thr}\to p\bar{p})<
1.6 \times 10^{-6}$ at 90$\%$ CL.

With a limited sample of 8.6 million $J/\psi$ available to us from
$\psi(2S)\to \pi^{+}\pi^{-}J/\psi$ we have searched for $J/\psi \to \gamma R_\mathrm{thr}$.
We find a $p\bar{p}$ threshold enhancement. When it is analyzed taking into account an
enhancement at $M(p\bar{p})=2100$ MeV/$c^{2}$, we obtain
$M(R_\mathrm{thr}) = 1837^{+10+9}_{-12-7}~\mathrm{MeV}/c^{2}$,
$\Gamma(R_\mathrm{thr}) = 0^{+44}_{-0}~ \mathrm{MeV}/c^{2}$, and
$\mathcal{B}(J/\psi \to \gamma R_\mathrm{thr}) \!\times\! \mathcal{B}(R_\mathrm{thr}\to p\bar{p}) = (11.4^{+4.3+4.2}_{-3.0-2.6}) \!\times\! 10^{-5}$.

\begin{table*}[tb]
\small
\begin{center}
\caption{Summary of the measured quantities. First errors are statistical,
second errors are systematic.}
\begin{tabular}{lccc}
\hline \hline
Quantity & Events & Our result & PDG08 \\
& & ($10^{-5}$) & ($10^{-5}$) \\
\hline
$\mathcal{B}(\psi(2S)\to \gamma p\bar{p})$ & $348\pm22$ & $4.18\pm0.26\pm0.18$ & $2.9\pm0.6$ \\
$\mathcal{B}(\psi(2S)\to \pi^{0} p\bar{p})$ & $948\pm37$ & $15.4\pm0.6\pm0.6$ & $13.3\pm1.7$ \\
$\mathcal{B}(\psi(2S)\to \eta p\bar{p})$ & $154\pm16$ & $5.6\pm0.6\pm0.3$ & $6.0\pm1.2$ \\
& & & \\
$\mathcal{B}(\psi(2S)\to \gamma f_{2}(1950))\times\mathcal{B}(f_{2}(1950)\to p\bar{p})$ & $111\pm19$ & $1.2\pm0.2\pm0.1$ & \\
$\mathcal{B}(\psi(2S)\to \gamma f_{2}(2150))\times\mathcal{B}(f_{2}(2150)\to p\bar{p})$ & $73\pm18$ & $0.72\pm0.18\pm0.03$ & \\
& & & \\
$\mathcal{B}(\psi(2S)\to \bar{p}N^{*}_{1}(1440))\times\mathcal{B}(N^{*}_{1}(1440)\to p\pi^{0})$ & $474\pm42$ & $8.1\pm0.7\pm0.3$ & \\
$\mathcal{B}(\psi(2S)\to \bar{p}N^{*}_{2}(2300))\times\mathcal{B}(N^{*}_{2}(2300)\to p\pi^{0})$ & $265\pm39$ & $4.0\pm0.6\pm0.2$ & \\
$\mathcal{B}(\psi(2S)\to \pi^{0}R_{1}(2100))\times\mathcal{B}(R_{1}(2100)\to p\bar{p})$ & $76\pm29$ & $1.1\pm0.4\pm0.1$ & \\
$\mathcal{B}(\psi(2S)\to \pi^{0}R_{2}(2900))\times\mathcal{B}(R_{2}(2900)\to p\bar{p})$ & $133\pm38$ & $2.3\pm0.7\pm0.1$ & \\
& & & \\
$\mathcal{B}(\psi(2S)\to \bar{p}N^{*}(1535))\times\mathcal{B}(N^{*}(1535)\to p\eta)$ & $123\pm16$ & $4.4\pm0.6\pm0.3$ & \\
$\mathcal{B}(\psi(2S)\to \eta R_{1}(2100))\times \mathcal{B}(R_{1}(2100)\to p\bar{p})$ & $31\pm10$ & $1.2\pm0.4\pm0.1$ & \\ \hline \hline
\end{tabular}
\end{center}
\normalsize
\end{table*}

\begin{acknowledgments}
We gratefully acknowledge the effort of the CESR staff
in providing us with excellent luminosity and running conditions.
D.~Cronin-Hennessy thanks the A.P.~Sloan Foundation.
This work was supported by
the National Science Foundation,
the U.S. Department of Energy,
the Natural Sciences and Engineering Research Council of Canada, and
the U.K. Science and Technology Facilities Council.
\end{acknowledgments}


\begin{thebibliography}{99}

\bibitem{prot1} J.-M. Richard, Nucl. Phys. Proc. Suppl. \textbf{86}, 361 (2000).

\bibitem{prot2} A. Antonelli \textit{et al.}, Nuclear Physics \textbf{B 517}, 3-35 (1998).

\bibitem{belle1} K. Abe \textit{et al.} (Belle Collaboration), Phys. Rev. Lett. 
\textbf{88}, 181803 (2002).

\bibitem{belle2} K. Abe \textit{et al.} (Belle Collaboration), Phys. Rev. Lett. 
\textbf{89}, 151802 (2002).

\bibitem{bes} J. Z. Bai \textit{et al.} (BES Collaboration), Phys. Rev. Lett. \textbf{91},
022001 (2003).

\bibitem{belle3} M.-Z. Wang \textit{et al.} (Belle Collaboration), Phys. Rev. Lett. 
\textbf{92}, 131801 (2004); Phys. Lett. \textbf{B 617}, 141 (2005).

\bibitem{babar1} B. Aubert \textit{et al.} (BaBar Collaboration), Phys. Rev. \textbf{D 72} 051101(R) (2005).

\bibitem{babar2} B. Aubert \textit{et al.} (BaBar Collaboration), Phys. Rev. \textbf{D 74} 051101(R) (2006).

\bibitem{belle4} J.-T. Wei \textit{et al.} (Belle Collaboration), Phys. Lett. \textbf{B 659}, 80 (2008).

\bibitem{threv} For a review of various suggestions see A. Sibirtsev \textit{et al.},
Phys. Rev. \textbf{D 71} 054010 (2005), and references therein.

\bibitem{cleodet} R. A. Briere \textit{et al.} (CLEO-c/CESR-c Taskforces $\&$ CLEO-c 
Collaboration), Cornell University LEPP Report No. CLNS 01/1742 (2001) unpublished; 
Y. Kubota \textit{et al.}, Nucl. Instrum. Methods Phys. Res., Sect. \textbf{A 320}, 66 (1992); 
D. Peterson \textit{et al.}, Nucl. Instrum. Methods Phys. Res., Sect. A \textbf{478}, 
142 (2002); 
M. Artuso \textit{et al.}, Nucl. Instrum. Methods Phys. Res., Sect. A \textbf{554}, 147 (2005).

\bibitem{jetset} T.~Sj\"ostrand \textit{et al.}, Comp. Phys. Commun. \textbf{135}, 238 (2001).

\bibitem{pdg08} C. Amsler \textit{et al.} (Particle Data Group), \textit{Phys. Lett.}
\textbf{B 667}, 1 (2008).

\bibitem{cleopp} P. Naik, \textit{et al.} (CLEO Collaboration), Phys. Rev. \textbf{D 78}, 031101(R) (2008).

\bibitem{crysball} S.B.~Athar \textit{et al}. (CLEO Collaboration), Phys. Rev. \textbf{D 70}, 112002 (2004).

\bibitem{besrad} M. Ablikim \textit{et al.} (BES Collaboration), Phys. Rev. Lett. \textbf{99}, 011802 (2007).

\bibitem{bespsip} M. Ablikim \textit{et al.} (BES Collaboration), Phys. Rev. \textbf{D 71}, 072006 (2005).

\bibitem{bes1837} M. Ablikim \textit{et al.} (BES Collaboration), Phys. Rev. Lett. \textbf{95}, 262001 (2005).

\bibitem{cleopsip} S. B. Athar \textit{et al.}, (CLEO Collaboration), Phys. Rev. \textbf{D 70}, 112002 (2004).

\bibitem{cleotrig} T. K. Pedlar \textit{et al.} (CLEO Collaboration), Phys. Rev. \textbf{D 72}, 051108(R) (2005).

\bibitem{cleogam} N. E. Adam \textit{et al.} (CLEO Collaboration), Phys. Rev. Lett. \textbf{94}, 012005 (2005).

\bibitem{cleotrack} Q. He \textit{et al.} (CLEO Collaboration), Phys. Rev. Lett. \textbf{95}, 121801 (2005).

\end{thebibliography}
\end{document}